\documentclass[journa,10pt,doublecolumn,draftcls]{IEEEtran}
\usepackage{amsmath,amsfonts}
\usepackage{algorithmic}
\usepackage{algorithm}
\usepackage{array}
\usepackage[caption=false,font=normalsize,labelfont=sf,textfont=sf]{subfig}
\usepackage{textcomp}
\usepackage{stfloats}
\usepackage{url}
\usepackage{verbatim}
\usepackage{graphicx}
\usepackage{cite}
\usepackage{braket}
\usepackage{amssymb}
\usepackage{yhmath}
\usepackage{color}
\usepackage{orcidlink}
\usepackage{geometry}
\usepackage{setspace}
\usepackage{orcidlink}
\usepackage{hyperref}
\hypersetup{hypertex=true,
	colorlinks=true,
	linkcolor=blue,
	anchorcolor=red,citecolor=blue}
\allowdisplaybreaks[2]

\begin{document}

\singlespacing

\title{Joint Beamforming Optimization for Active STAR-RIS-Assisted ISAC Systems}

\author{Shuang Zhang, Wanming Hao,~\IEEEmembership{Senior Member,~IEEE,} Gangcan Sun, Chongwen Huang,~\IEEEmembership{Member,~IEEE,} Zhengyu Zhu,~\IEEEmembership{Senior Member,~IEEE,}  Xingwang Li,~\IEEEmembership{Senior Member,~IEEE,} Chau Yuen,~\IEEEmembership{Fellow,~IEEE}
	\thanks{This work was supported by the China National Natural Science Foundation under Grant 62101499, the open research fund of National Mobile Communications Research Laboratory Southeast University under grant 2024D12 and 2024D15, and the Program for Science \& Technology Innovation Talents in Universities of Henan Province under Grant 24HASTIT038. The work of C. Huang was supported by the China National Key R\&D Program under Grant 2021YFA1000500 and 2023YFB2904800, and the National Natural Science Foundation of China under Grant 62101492. The work of X. Li was supported by the Key Research and Development Project of Henan Province under Grant 231111210500. The work of C. Yuan was supported by the Ministry of Education Singapore MOE Tier 2 (Award number MOET2EP50220-0019). (\emph{Corresponding author: Wanming Hao})}
	\thanks{S. Zhang, W. Hao, G. Sun, Z. Zhu are with the School of Electrical and Information Engineering, Zhengzhou University, Zhengzhou 450001, China, and W. Hao is also with the National Mobile Communications Research Laboratory, Southeast University, Nanjing 214135, China (E-mail: yayue96@163.com, \{iewmhao,iegcsun,iezyzhu\}@zzu.edu.cn).}
		\thanks{C. Huang is with College of Information Science and Electronic Engineering, Zhejiang University, Hangzhou 310027, China, and Zhejiang Provincial Key Laboratory of Info. Proc., Commun. \& Netw. (IPCAN), Hangzhou 310027, China (E-mail: chongwenhuang@zju.edu.cn).}
	\thanks{X. Li is with the School of Physics and Electronic Information Engineering, Henan Polytechnic University, Jiaozuo, China, and also with the National Mobile Communications Research Laboratory, Southeast University, Nanjing 214135, China (email: lixingwang@hpu.edu.cn).}
	\thanks{C. Yuen is with the School of Electrical and Electronics Engineering, Nanyang Technological University, 639798, Singapore (e-mail: chau.yuen@ntu.edu.sg).}}

\maketitle

\begin{abstract}
In this paper, we investigate an active simultaneously transmitting and reﬂecting reconfigurable intelligent surface (STAR-RIS) assisted integrated sensing and communications (ISAC) system, where the dual-function base station (DFBS) operates in full-duplex (FD) mode to provide communication services and performs targets sensing simultaneously. Meanwhile, we consider multiple targets and multiple users scenario as well as the self-interference at the FD DFBS. Through jointly optimizing the DFBS and active STAR-RIS beamforming under different work modes, our purpose is to achieve the maximum communication sum-rate, while satisfying the minimum radar signal-to-interference-plus-noise ratio (SINR) constraint, the active STAR-RIS hardware constraints and the total power constraint of DFBS and active STAR-RIS.
To tackle the complex non-convex optimization problem formulated, an efficient alternating optimization algorithm is proposed.
Specifically, the fractional programming method is first leveraged to turn the original problem into a more tractable one, and subsequently the transformed problem is decomposed into several sub-problems. Next, we develop a derivation method to obtain the closed-form expression of the radar receiving beamforming, and then the DFBS transmit beamforming is optimized under the radar SINR requirement and total power constraints. After that, the active STAR-RIS reflection and transmission beamforming are optimized by majorization minimization, complex circle manifold and convex optimization techniques. Finally, the proposed schemes are conducted through numerical simulations to show their benefits and efficiency. 
\end{abstract}


\section{Introduction}
\IEEEPARstart{R}{ecently,} with the development of radar and communication technologies, their differences in antenna structure, hardware component architecture and working bandwidth have been gradually reduced, which attracts widespread attention on integrated sensing and communication (ISAC) in academic and industrial fields \cite{ref1,ref2}. By the utilization of shared hardware platforms, spectrum and signal processing frameworks, ISAC can integrate sensing and communication tightly and seek their performance balance, thereby significantly enhancing the spectral efficiency, decreasing the hardware cost and system power consumption, which has made it gain recognition as a promising technology for the forthcoming sixth generation (6G) wireless communication \cite{ref3}. However, ISAC still faces several challenges in practical applications, such as high environmental dependence, limited coverage, and many blind spots. Fortunately, owing to the ability to create efficient line-of-sight (LOS) links and provide additional degrees of freedom (DoFs) by smartly manipulating the wireless propagation environment, the reconfigurable intelligent surface (RIS) offers a novel approach to meet these challenges \cite{ref4,ref5,ref6,ref7}. Typically, RIS consists of numerous reconfigurable elements with low power and low cost, each of which can dynamically adjust the phase shift and amplitude of the incident signal, allowing for a controllable propagation environment\cite{ref8,ref9}. Deploying RIS can provide a more controllable propagation environment, improving the coverage range and transmission quality \cite{ref10,ref11,ref12,ref13}.
\IEEEpubidadjcol
\subsection{Related works}
Nowadays, there have been several works considering integrating RIS into ISAC systems \cite{ref14,ref15,ref16,ref17,ref18,ref19,ref20,ref21,ref22}, in which the joint beamforming optimization of RIS and base station (BS) is proposed to improve the system performance. To be specific, in \cite{ref14}, a RIS-assisted dual-function radar and communication system was designed for the first time with single-user and single-target scenario. Then, the radar signal-to-noise ratio (SNR) maximization problem was formulated under transmit power budget and quality of service (QoS) constraint, and an iterative optimization algorithm was introduced based on majorization minimization (MM), semidefinite programming (SDP) and semidefinite relaxation (SDR) techniques to solve it. The authors \cite{ref15} extended their work to the multi-user scenario, where the RIS reflection coefficients, BS transmit beamforming and receive filter were jointly designed to maximize the sum-rate. To address the optimization problem, an iterative algorithm framework was introduced, utilizing the alternative direction method of multipliers (ADMM), fractional programming (FP) and MM techniques. In \cite{ref16}, the authors considered the interference of strong multiple clutters on target detection, and aimed to maximize the radar signal-to-interference-plus-noise ratio (SINR) while ensuring the QoS requirements for communication, while ensuring the QoS requirements for communication. To deal with it, an algorithm framework based on ADMM and MM was introduced. In \cite{ref17}, the authors utilized RIS to assist the multiple targets detection, and a minimum gain of RIS’s \textcolor{blue}{beampattern} effective algorithm framework was exploited to obtain the maximized expected detection angles, while meeting the minimum communication SNR constraint. The authors \cite{ref18} further considered a more general scenario, where the BS performed multiple users communication and multiple targets sensing. The weighted summation of target detection SNRs was maximized by comprehensively applying penalty-based, manifold optimization and MM methods. In \cite{ref19} and \cite{ref20}, a new joint optimization metric was introduced. Specifically, work \cite{ref19} attempted to utilize the joint optimization metric to maximize the radar SINR, and meanwhile minimizing the multi-user interferences. In work \cite{ref20}, the cross-correlation pattern, beam pattern error and total interference between multiple users and radar were minimized. The work \cite{ref21} concentrated on minimizing the BS transmit power by taking into account the interference caused by the cross-correlation between radar and RIS, and an alternating optimization (AO) algorithm based on SDR was proposed. The authors \cite{ref22} created a novel strategy of RIS-assisted beamforming in consideration of the dimensions of the target, where the conception of ultimate detection resolution (UDR) based on target size was introduced to evaluate the capability of target detection the first time. Then an optimization problem of maximizing user SNR was formulated subject to the minimum UDR requirement.\\
\indent However, the aforementioned researches assume that RIS only reflects the incident signals. In this case, it is necessary for the communication users or targets to be placed on the same side of RIS and BS, resulting in a coverage limited to half-space. Given that the spatial confinement may not always be satisfied, it significantly restricts the flexibility and efficacy of RIS. Recently, there has been a growing interest in a burgeoning concept known as intelligent omni-surface (IOS) or simultaneously transmitting and reflecting (STAR)-RIS \cite{ref23,ref24,ref25,ref26}, which can achieve full-space coverage. A STAR-RIS-enabled ISAC framework was first brought up in \cite{ref27}. To overcome the severe path loss and interference from clutter in sensing, a new type of STAR-RIS sensor structure was introduced, where the dedicated sensors were installed at the STAR-RIS. Furthermore, to explore the ability of the typical STAR-RIS in ISAC systems, the authors \cite{ref28} considered an ISAC network enhanced by energy-splitting STAR-RIS. Under the objective of maximizing sensing SNR, an iterative algorithm was created based on MM, sequential rank-one constraint relaxation and SDR techniques. After that, the authors \cite{ref29} undertook the initial exploration of leveraging STAR-RIS to enhance the security efficacy of ISAC systems, and a secrecy rate maximization problem was formulated while ensuring the sensing minimum SNR. The authors \cite{ref30} further considered the physical layer security of ISAC systems with the presence of multiple eavesdropping targets. To maximize the average received radar sensing power, a low-complexity approach based on distance-majorization was conducted. As a further advance, the authors \cite{ref31} utilized the long-term synergistic effect between STAR-RIS and ISAC and the two deep reinforcement learning algorithms to augment legitimate users' mean long-term security rate.\\ 
\indent Nevertheless, the above works in ISAC systems mainly focus on the passive STAR-RIS. Although the ability to simultaneously reflect and transmit signals enables passive STAR-RIS to realize full-space coverage, it still has the inherent flaws of passive RIS, namely multiplicative fading \cite{ref32}. This severe twin-hop path-loss fading will limit the performance improvement of passive STAR-RIS. Therefore, inspired by the active RIS introduced \cite{ref33,ref34,ref35}, to overcome the multiplicative fading and significantly improve space coverage, a new concept has recently been proposed, namely active STAR-RIS. The advantage of active STAR-RIS is that it can not only reflect and transmit the incident signals, but also amplify the incident signal to compensate for the multiplicative fading. Hence, it can achieve the same performance as passive STAR-RIS, but requires much fewer elements, which means that the complexity of channel estimation and beamforming can be significantly reduced. Compared to the performance gain attained by enlarging the STAR-RIS size, the gain brought by active nature is more appealing and practical. \\
\indent Currently, there have been some preliminary studies on active STAR-RIS, including \cite{ref36,ref37,ref38,ref39,ref40,ref41,ref42}. The concept of active STAR-RIS was first brought up in \cite{ref36} to assist vehicular communication system, where the transmit precoding matrix at the BS and active STAR-RIS coefficient matrices were jointly optimized to minimize the BS’s transmit power. The authors \cite{ref37} introduced active STAR-RIS into the non-orthogonal multiple access system, where the beamforming vector at the BS and the active STAR-RIS coefficients were jointly optimized to obtain the maximized sum achievable rate. The authors \cite{ref38} considered the joint BS transmit beamforming and active STAR-RIS coefficients design in multiuser multiple-input single-output systems, in which both sum-rate maximization and power minimization problems were studied. The authors \cite{ref39} proposed an efficient beamforming scheme for active STAR-RIS-aided wireless communications, where the sum-rate was maximized. The authors \cite{ref40} investigated the problem of maximizing the secrecy rate underlaying an active STAR-RIS-enabled communication system, where the access point’s beamforming and active STAR-RIS configuration were jointly designed. The authors \cite{ref41} studied the joint design of transmit beamforming and the active STAR-RIS configuration in a multi-user MIMO system. With the purpose of maximizing spectral efficiency, an analytic-based, highly computation-parallelizable, and convergence-guaranteed algorithm was developed. The authors \cite{ref42} deployed an active STAR-RIS in a NOMA downlink multiuser network. By jointly optimizing the active beamforming and active STAR-RIS coefficients, the sum-rate was maximized.

\subsection{Main Contributions}

Based on the above analysis, the recent works of active STAR-RIS focused on its application in communication systems. To the best of our knowledge, the utilization of active STAR-RIS to ISAC systems has yet to be investigated. Especially, the joint beamforming design of active STAR-RIS and DFBS transceiver, the multiple targets and multiple users scenario, as well as the self-interference (SI) introduced by DFSB full-duplex (FD) mode, all bring challenges for the application of active STAR-RIS in ISAC systems. Thus, motivated by the above discussions, we investigate the joint design and optimization of active STAR-RIS-assisted ISAC systems to fill the research gap. Concretely, we propose an ISAC system empowered by active STAR-RIS, wherein the DFBS operates in FD mode and transmits the dual-functional signals, thereby providing communication services for multiple user with the help of an active STAR-RIS and performing multiple targets sensing concurrently. For the proposed system, the joint beamforming design of the DFBS and active STAR-RIS is examined to maximize the communication sum-rate, while ensuring the SINR of all targets surpasses a predefined threshold. Overall, the main contributions of our work can be summarized as follows:

\begin{itemize}
\item{We investigate an active STAR-RIS-assisted ISAC system for mitigating the multiplicative fading effect as well as enhancing full-space coverage. This system leverages a DFBS and an active STAR-RIS to serve multiple users and detect multiple targets. Meanwhile, we consider the existing three work modes of active STAR-RIS, i.e., unequal energy division (UED), equal energy division (EED) and space division (SD) modes. In order to enhance practicality, we consider the FD mode and SI at the DFBS, as well as the echo interferences caused by multiple targets.}

\item{On this basis, we formulate an optimization problem, which aims to maximize the communication sum-rate through the joint design of DFBS transmit/receive beamforming and active STAR-RIS beamforming, subject to the minimum sensing SINR constraint per target, the active STAR-RIS hardware constraints and the transmit power constraints at DFBS and active STAR-RIS. Due to the high coupling of variables, the objective function is non-convex with fractional form, as well as the sensing SINR constraint and active STAR-RIS phase shift constraint are both non-convex, so it poses a significant challenge to solve the formulated problem directly. Thus, the FP method is leveraged first to convert the original problem into a more tractable structure using the quadratic transform and the Lagrangian dual transform. Next, based on AO framework, the transformed problem is decomposed into three subproblems, which can be iteratively optimized.}

\item{Then, we offer a derivation method for the closed expression of the radar receiving beamforming by reducing it into a typical generalized “Rayleigh quotient” optimization problem. Subsequently, the DFBS transmit beamforming can be transformed into a conventional quadratic constraint quadratic programming (QCQP) optimization problem that can be solved with standard convex optimization algorithms. In the end, the active STAR-RIS reflection/transmission beamforming is designed under three active STAR-RIS work modes. For UED mode, with the other variables given, the subproblem can also be reformulated as a QCQP optimization problem, and the standard convex optimization algorithm can be applied to obtain the solution. For EED and SD modes, the subproblem can be addressed by complex circle manifold and MM techniques.}

\item{Finally, numerical simulation results demonstrate the convergence and effectiveness of our proposed algorithms. Our results show that the proposed active STAR-RIS schemes obtain a higher performance compared to the passive STAR-RIS, which validates the benefits of deploying active STAR-RIS in ISAC systems.}
\end{itemize}

The rest of this paper is organized as follows. In Section II, the system model of the downlink active STAR-RIS-assisted ISAC system is introduced, and the maximum communication sum-rate optimization problem is formulated. Section III presents the proposed joint beamforming optimization framework under three active STAR-RIS work modes. The simulation results are shown in Section IV. Finally, conclusions are given in Section V.
\begin{table}[!t]
	\caption{Notations \label{tab:table1}}
	\centering
	\resizebox{0.5\textwidth}{!}{
		\begin{tabular}{|m{1.7cm}<{\centering}||m{7.6cm}<{\centering}|}
			\hline
			\textbf{Notation} & \textbf{Description}\\
			\hline
			$M$ & The number of DFBS transmit/receive antennas\\
			\hline
			$N$ & The number of active STAR-RIS elements\\ 
			\hline
			$K$/$Q$ & The total number of users and targets\\
			\hline
			$K_r$/$K_t$ & The number of users at reflection space and transmission space\\
			\hline
			$\boldsymbol{\Psi}_{r}$/ $\boldsymbol{\Psi}_{t}$ & The reflection and transmission coefficient matrices of active STAR-RIS\\
			\hline
			$\boldsymbol{\Phi}_{r}$/$\boldsymbol{\Phi}_{t}$ & The reflection and transmission phase-shift matrices of active STAR-RIS\\
			\hline
			$\mathbf{A}_{r}$/$\mathbf{A}_{t}$ & The reflection and transmission amplitude matrices of active STAR-RIS\\
			\hline
			$\boldsymbol{\psi}_{r}$/$\boldsymbol{\phi}_{r}$/$\boldsymbol{a}_{r}$ & Equivalent active STAR-RIS reflection coefficient, phase-shift and amplitude vectors\\
			\hline
			$\boldsymbol{\psi}_{t}$/$\boldsymbol{\phi}_{t}$/$\boldsymbol{a}_{t}$ & Equivalent active STAR-RIS transmission coefficient, phase-shift and amplitude vectors\\
			\hline
			$\textbf{x}$ & Transmission signal at the DFBS \\
			\hline
			$\textbf{s}_c$/$\textbf{s}_r$ & Communication signal and radar signal\\
			\hline
			$\textbf{W}_c$/$\textbf{W}_r$ & Beamforming matrix for users and targets\\
			\hline
			$\textbf{W}$/$\textbf{s}$ & Equivalent beamforming matrix and transmission signal at the DFBS \\
			\hline
			$\textbf{w}_j$ & Equivalent beamforming vector at $\textbf{W}$ $j$-th column\\
			\hline
			$\textbf{G}$ & Channel from the DFBS to the active STAR-RIS \\
			\hline
			$\textbf{f}_k$ & Channel from the active STAR-RIS to the $k$-th user\\
			\hline
			$\textbf{h}_{d,k}$/$\textbf{h}_{d,q}$ & Direct channels from the DFBS to the $k$-the user and $q$-th target \\
			\hline
			$\tilde{\textbf{h}}_{r,k}^{H}$/$\tilde{\textbf{h}}_{t,k}^{H}$ & Equivalent cascaded channels from the DFBS to the $k$-th user \\
			\hline
			$\textbf{H}_{SI}$ & SI channel between the DFBS transmitter and receiver \\
			\hline
			$b$/$\xi^2$ & The target RCS and the corresponding variance \\
			\hline
			$\textbf{v}$/$\textbf{z}_q$/$n_k$ & Thermal noise of active STAR-RIS and complex AWGN of the $q$-th target and the $k$-th user \\
			\hline
			$\sigma^2_v$/$\sigma^2_z$/$\sigma^2_k$ & Variances of $\textbf{v}$, $\textbf{z}_q$ and $n_k$ \\
			\hline
			$\textbf{u}_q$ & The radar receive filter for $q$-th target \\
			\hline
			$\Gamma_q$ & The sensing SINR requirement for the $q$-th target \\
			\hline
			$P_{\text{max}}^{\text{B}}$/$P_{\text{max}}^{\text{R}}$ & The power budget at the DFBS and active
			STAR-RIS \\
			\hline
		\end{tabular}
	}
\end{table}

\textit{Notations}: Throughout this paper, the following notations are employed. Column vectors are typically indicated by boldface lower-case letters, whereas matrices are commonly represented by boldface upper-case letters.
$(\cdot)^{T}$, $(\cdot)^{*}$, $(\cdot)^{H}$, and $(\cdot)^{-1}$ indicate the transpose, conjugate, conjugate-transpose, and the inversion operations of a matrix, respectively. 
$|\cdot|$ and $\Re\{\cdot\}$ refer to the modulus and real part of a complex variable, respectively. 
$\mathbb{E\{\cdot\}}$ and $\mathbb{C}^{m\times n}$ denote statistical expectation vaule and the complex space with a dimension of $m\times n$, respectively. 
diag$(\cdot)$ refers to create a diagonal matrix from a vector or the diagonal elements of a matrix. 
$\mathbf{I}_{M}$ indicates the $M\times M$ identity matrix. 
$\otimes $ refers to the Kronecker product. 
$\|\cdot\|_{F}$ indicates the F-norm of the input matrix, and arg$(\cdot)$ is the phase of complex variable. Besides, we have summarized the main symbols and their corresponding means in TABLE I to provide a better reading experience and facilitate understanding.
\begin{figure*}[!t]
	\centering
		\subfloat[]{\includegraphics[height=2.2in,width=3in]{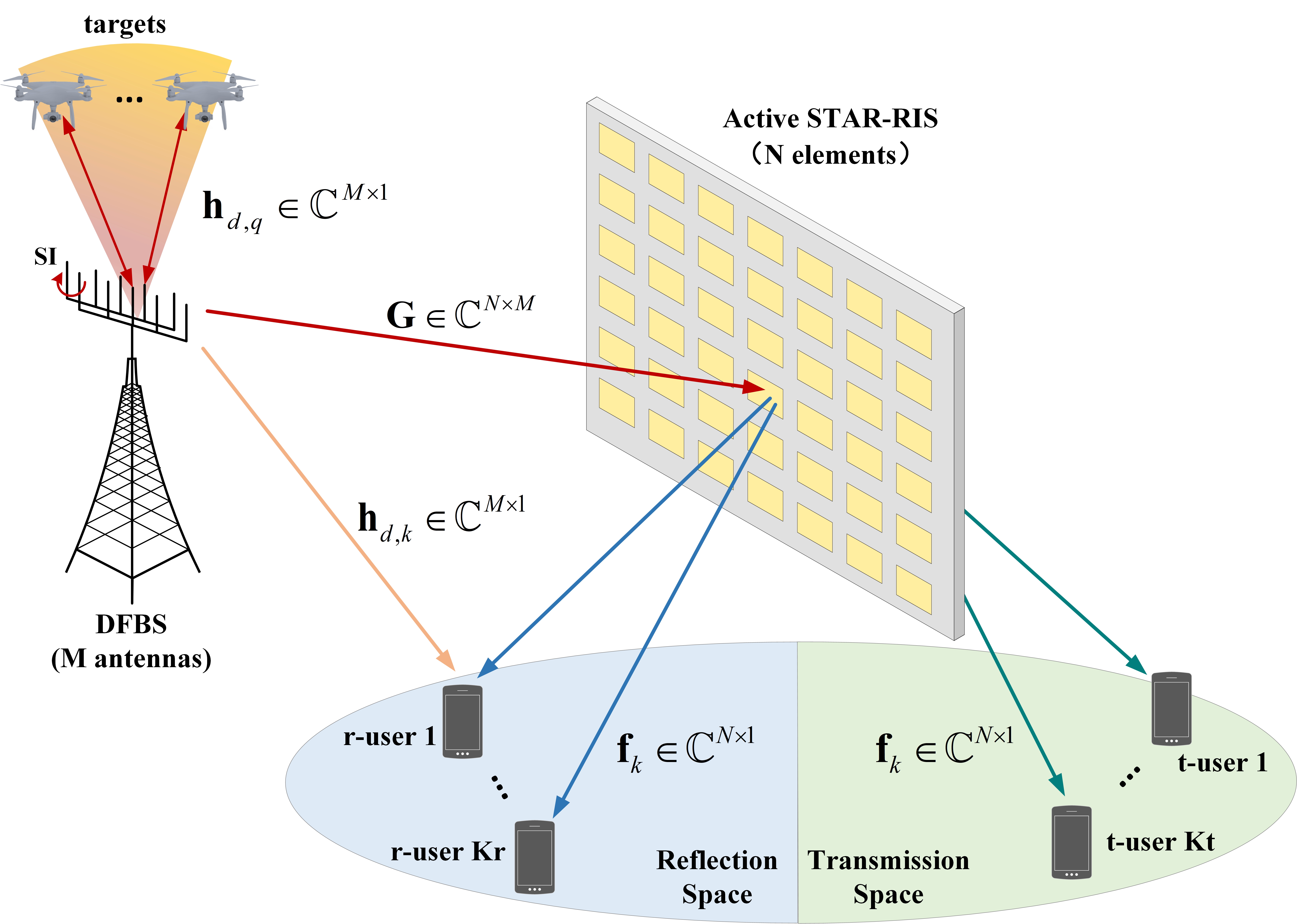}%
			\label{(fig_first_case)}}
		\hfil
		\subfloat[]{\includegraphics[height=2.45in,width=3.9in]{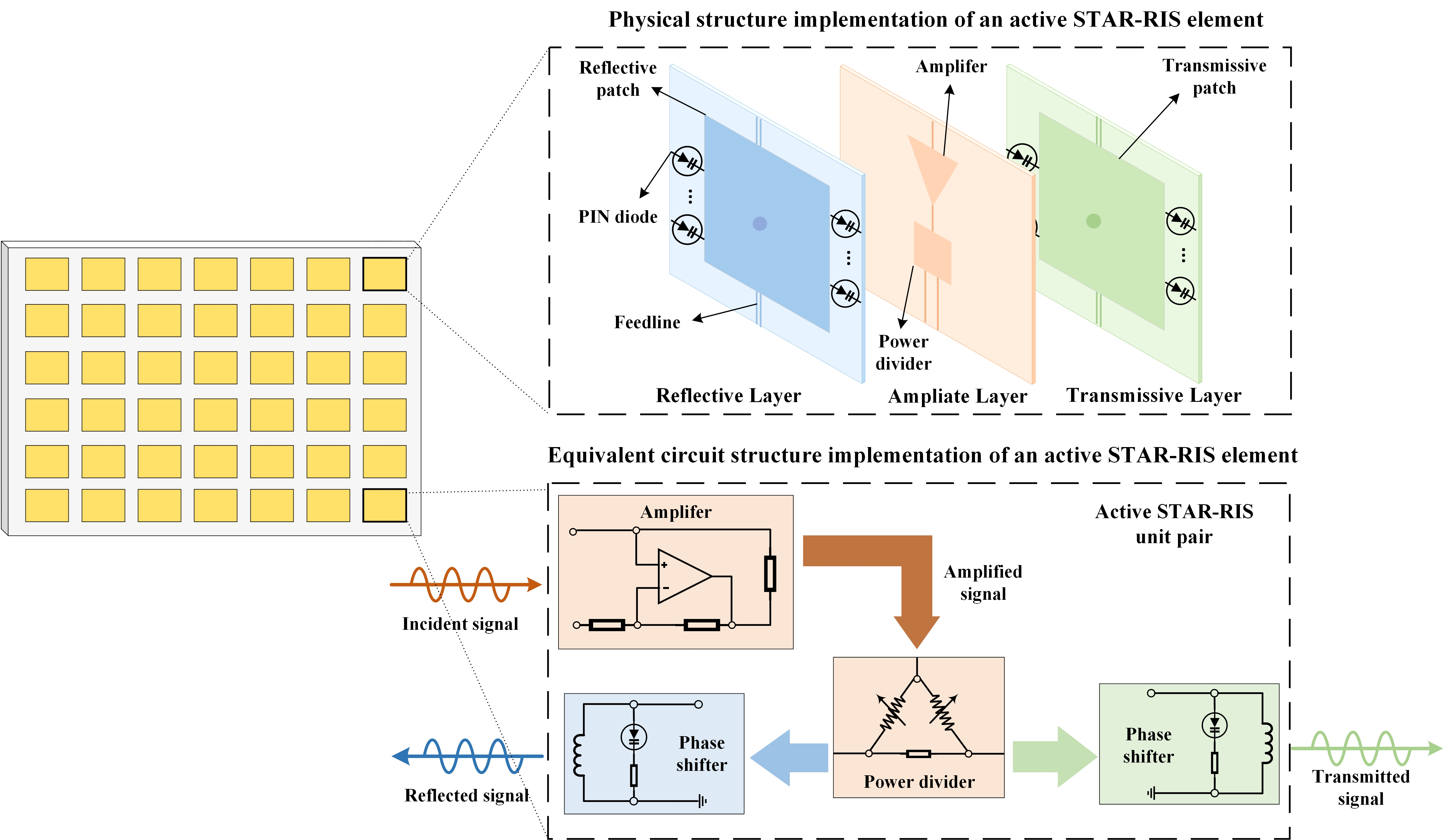}%
			\label{fig_second_case}}
		\caption{Structure of active STAR-RIS and its deployment in a ISAC system. (a) Active STAR-RIS-assisted ISAC system model. (b) The basic structure of the active STAR-RIS.}
		\label{fig_1}
\end{figure*}

\section{System Model and Problem Formulation}
An active STAR-RIS-assisted ISAC system is considered as depicted in Fig. \ref{fig_1}(a), in which a DFBS with $M$ transmit/receive antennas provides communication services to $K$ single-antenna users with the assistance of an $N$-element active STAR-RIS and performs $Q$ targets sensing simultaneously. Here, we define the users located in the reflection space as the $r$-user and can be denoted by the set of $\mathcal K_{r}\triangleq\{1,2,\dots,K_{r}\}$, while the users located in the transmission space is the $t$-user and can be denoted by the set of $\mathcal K_{t}\triangleq\{1,2,\dots,K_{t}\}$, and we have $K_{r}+K_{t}=K$. The active STAR-RIS is assumed to be deployed in close proximity to the users for assisting downlink communications effectively. In the meantime, we assume that the targets fly at low altitudes and own the robust LoS link with the DFBS. Considering the round-trip characteristic of the echo signals, the signals that go through the DFBS-target-active STAR-RIS-DFBS channel are highly attenuated due to three-hop transmissions, and since the active STAR-RIS is far from the targets, which will cause the three-hop path loss to be more severe. Thus, compared to the echoes reflected by STAR-RIS and undergoing three-hop attenuation to reach DFBS, the target echoes will be much larger\footnote[1]{Deploying active STAR-RIS is used to assist DFBS to communicate with users for improving the service of quality. Therefore, the active STAR-RIS reflect/transmit beamforming formulated by the proposed scheme will toward to the r-/t-users, rather than reflecting back to DFBS. In this case, even the DFBS receives the echo from the active STAR-RIS, the received echo power from DFBD-active STAR-RIS-DFBS link is very low and can be almost negligible.}. Therefore, similar to the prior works \cite{ref21}, \cite{ref28}, \cite{ref43, ref44}, the reflected clutter interferences through active STAR-RIS in our paper are omitted due to the severe round-trip path loss and physical obstacles.

\subsection{Active STAR-RIS And Signal Models}
The active STAR-RIS owns the ability to amplify, transmit, and reflect signals simultaneously via amplifiers and phase shifters. Each element of the active STAR-RIS can have the incident signals' amplitude and phase-shift adjusted separately using different components \cite{ref36,ref37,ref38,ref39,ref40,ref41,ref42}. As shown in the upper half of Fig. \ref{fig_1}(b), the physical structure of each active STAR-RIS element consists of a reflective layer, an ampliate layer, and a transmissive layer. The reflective layer and transmissive layer are two symmetric layers, each one contains a metal patch, multiple positive intrinsic negative (PIN) diodes, and a feedline. By controlling the bias voltage of PIN diodes through the feedline, they can be work on the ON or OFF state so as to realize the different phase shifts, and thus the phase shift can be adjusted flexible. The power amplifier and power divider embedded in the ampliate layer can be effectively designed to change the amplitude of reflected and transmitted signals \cite{ref38,ref42}. The equivalent circuit implementation can be found in the lower half of Fig. \ref{fig_1}(b), the active STAR-RIS structure deploys element arrays on two opposite faces of the plate with the element aligned. Each pair of aligned elements on the opposite faces are connected by circuits embedded within the plate. The incident signals on the active STAR-RIS are first fed into the reflection-type amplifier (RA) for amplification. Then, the amplified signals are divided into two parts by the power divider (PD), and are fed into a pair of opposite controllable phase shifters in the reflective and transmissive layers. The signals are then emitted from the front and back sides of the active STAR-RIS and represented as reflected signals and transmitted signals, respectively, thus achieving simultaneous reflection and transmission of the incident signals. It is worth noting that the active STAR-RIS structure is indeed feasible and can be implemented based on existing technologies\footnote[2]{The key components of the active STAR-RIS are phase shifters, RA, and PD. For phase shifters, they can be implemented by the phase shifters that widely used in passive RIS/STAR-RIS. The implementation of the RA and PD, can refer to various existing analog circuit technologies \cite{ref38,ref40,ref41}. Specifically, RA can be implemented using various analog circuits, such as, the aperture-coupled microstrip paths, tunnel diodes, field effect transistors, and even some complementary metal oxide semiconductors. PD can also be implemented utilizing low-cost analogue components, such as combining adjustable varactors/diodes with conventional power dividers or couplers, reconfigurable synthesized transmission line techniques, and the new Type-II reconfigurable couplers.}. 

Hence, based on the above analysis, the reflection and transmission matrices for active STAR-RIS are defined as $\boldsymbol{\Psi}_{r}\triangleq\mathbf{A}_r\boldsymbol{\Phi}_{r}=$ diag$(\boldsymbol{\psi}_{r})\in\mathbb{C}^{N\times N}$ and $\boldsymbol{\Psi}_{t}\triangleq\mathbf{A}_t\boldsymbol{\Phi}_{t}=$ diag$(\boldsymbol{\psi}_{t})\in\mathbb{C}^{N\times N}$, respectively, where $\boldsymbol{\Phi}_{r}\triangleq$ diag$(\boldsymbol{\phi}_{r})=$ diag$([e^{j\varphi_{r,1}},\dots,e^{j\varphi_{r,N}}]^{T})\in\mathbb{C}^{N\times N}$ denotes the reflection phase-shift matrix, $\boldsymbol{\Phi}_{t}\triangleq$ diag$(\boldsymbol{\phi}_{t})=$ diag$([e^{j\varphi_{t,1}},\dots,e^{j\varphi_{t,N}}]^{T})\in\mathbb{C}^{N\times N}$ denotes the transmission phase-shift matrix, 
$\mathbf{A}_{r}\triangleq$ diag$(\boldsymbol{a}_{r})=$ diag$([a_{r,1},\dots,a_{r,N}]^{T})$ denotes the reflection amplitude matrix, 
and $\mathbf{A}_{t}\triangleq$ diag$(\boldsymbol{a}_{t})=$ diag$([a_{t,1},\dots,a_{t,N}]^{T})$ denotes the transmission amplitude matrix.
	\begin{figure}[!t]
	\centering
	\includegraphics[height=1.65in,width=3.5in]{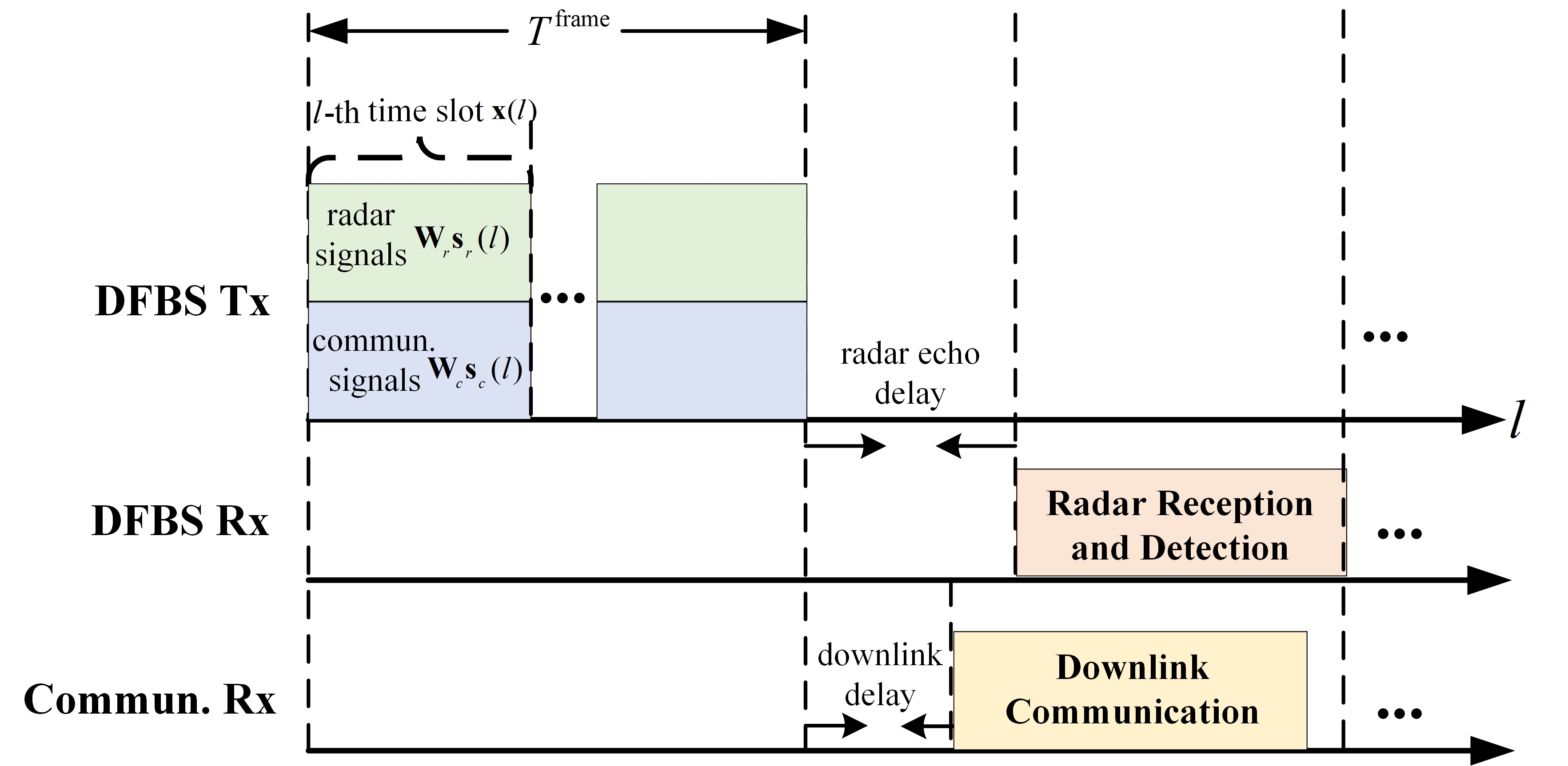}
	\caption{The dual-functional signal transmission diagram.}
	\label{fig_2}
\end{figure} 

The DFBS transmit a dual-functional signal that includes radar signals and communication signals. As illustrated in Fig. \ref{fig_2}, we fix our attention on one typical transmission frame, which comprises several time slots, thus, the transmitted signal in the $l$-th time slot is given by:
\begin{equation}\label{eqn_1}
	\textbf{x}(l)=\textbf{W}_{c}\textbf{s}_{c}(l)+\textbf{W}_{r}\textbf{s}_{r}(l)
		=\textbf{W}\textbf{s}(l),
\end{equation}
where $\textbf{s}_c(l)=\left[s_{c, 1}(l), \ldots, s_{c, K}(l)\right]^T \in \mathbb{C}^{K \times 1} \quad$ and $\textbf{s}_r(l)=\left[s_{r, 1}(l), \ldots, s_{r, M}(l)\right]^T \in \mathbb{C}^{M \times 1}$ denote the communication signals and radar signals, respectively, $\textbf{W}_c=\left[\textbf{w}_{c, 1}, \ldots, \textbf{w}_{c, K}\right] \in \mathbb{C}^{M \times K} \quad$ and $\quad \textbf{W}_r=\left[\textbf{w}_{r, 1}, \ldots, \textbf{w}_{r, M}\right] \in \mathbb{C}^{M \times M}$ are their corresponding beamformers with $\textbf{w}_{c, i}$ or $\textbf{w}_{r, i} \in \mathbb{C}^{M \times 1}$. To avoid mutual interference in radar and communication, the radar signals and communication signals are assumed to be independent of each other and own unit power, i.e., $\mathbb{E}\left\{\textbf{s}_c(l) \textbf{s}_r^H(l)\right\}=\mathbf{0}$, $\mathbb{E}\left\{\textbf{s}_c(l) \textbf{s}_c^H(l)\right\}=\mathbf{I}_K$ and $\mathbb{E}\left\{\textbf{s}_r(l) \textbf{s}_r^H(l)\right\}=\mathbf{I}_M$. We also assume that the communication signals for different users are uncorrelated, i.e., $\mathbb{E}\left\{s_{c, i}(l) s_{c, j}^*(l)\right\}=0, \forall i \neq j$. For simplicity, in \eqref{eqn_1}, we define $\textbf{W} \triangleq\left[\textbf{W}_c, \textbf{W}_r\right] \in \mathbb{C}^{M \times(K+M)}$ with $\textbf{w}_j \in \mathbb{C}^{M \times 1}$ denotes the $j$-th column of $\textbf{W}$, $j=1,2,...,(K+M)$, and $\textbf{s}(l) \triangleq\left[\textbf{s}_c^T(l), \textbf{s}_r^T(l)\right]^T \in \mathbb{C}^{(K+M) \times 1}$. As our proposed ISAC system needs to serve multi-user communication as well as target	sensing, the communication and sensing operations are synchronized such that the transmit antennas emit the dual-functional signals, while the receive antennas receive the echoes from the targets during the downlink communication. Note that since the channel state information remains unchanged in one frame, and the operations are the same for each time slot, without	loss of generality and for notational convenience, we omit the time index in the subsequent description of our paper.
\\ \indent For downlink communication, let $\textbf{G}\in\mathbb{C}^{N\times M}$, $\textbf{f}_{k}\in\mathbb{C}^{N\times 1}$ and $\textbf{h}_{d,k}\in\mathbb{C}^{M\times 1}$ 
represent the channels from the DFBS to the active STAR-RIS, from the active STAR-RIS to the $k$-th user and from the DFBS to the $k$-th user, respectively. In order to fully exploit the potential of active STAR-RIS, we assume that it is available for DFBS to obtain the accurate and complete channel state information. Thus, for $r$-user, the received signal at the $k$-th user can be expressed as:
\begin{equation}
	\label{eqn_2}
	\begin{split}
		y_{k} &=\textbf{h}^H_{d,k}\textbf{x}+\textbf{f}^H_{k}\boldsymbol{\Psi}_{r}(\textbf{G}\textbf{x}+\textbf{v})+n_{k}
		\\
		&=(\textbf{h}^{H}_{d,k}+\textbf{f}^H_{k}\boldsymbol{\Psi}_{r}\textbf{G})\textbf{W}\textbf{s}+\textbf{f}^H_{k}\boldsymbol{\Psi}_{r}\textbf{v}+n_{k}, \enspace k\in\mathcal K_{r},
	\end{split}
\end{equation}
where $\mathbf{v} \in \mathbb{C}^{N \times 1}$ is the thermal noise that related to the inherent device noise of the STAR-RIS elements with each entry obeying $\mathcal{C N}\left(0, \sigma_v^2\right)$, and $\boldsymbol{\Psi}_{r}\textbf{v}$ is the dynamic noise introduced by the amplifier in active STAR-RIS, also known as the amplified thermal noise. $n_{k}\sim \mathcal C\mathcal N(0,\sigma^2_{k})$ represents the complex additive white Gaussian noise (AWGN) at the $k$-th user with variance $\sigma^2_{k}$. Due to the use of active components, active STAR-RIS consumes additional power to amplify the incident signals, and the dynamic noise in active STAR-RIS cannot be neglected. Thus, SINR of the $k$-th $r$-user can be calculated as:
 \begin{equation}
 	\label{eqn_3}
 	\text{SINR}_{k}\!=\frac{|\tilde{\textbf{h}}^{H}_{r,k}\textbf{w}_{k}|^2}{\sum\limits^{K+M}\limits_{i=1,i\neq k}|\tilde{\textbf{h}}^{H}_{r,k}\textbf{w}_{i}|^2\!+\!\sigma^2_{v}\|\textbf{f}^{H}_{k}\boldsymbol{\Psi}_{r}\|^2\!+\!\sigma^2_{k}}, \; k\!\in\!\mathcal K_{r},
 \end{equation}

\noindent where $\tilde{\textbf{h}}_{r,k}^{H}\triangleq \textbf{h}^{H}_{d,k}+\textbf{f}^{H}_{k}\boldsymbol{\Psi}_{r}\textbf{G}\in\mathbb{C}^{1\times M}$ is the equivalent cascaded channels from the DFBS to the $k$-th $r$-user.
\\ \indent Similarly, the received signal for the $k$-th $t$-user can be calculated as:
\begin{equation}
	\label{eqn_4}
	\begin{split}
		y_{k} &=\textbf{h}^H_{d,k}\textbf{x}+\textbf{f}^H_{k}\boldsymbol{\Psi}_{t}(\textbf{G}\textbf{x}+\textbf{v})+n_{k}
		\\
		&=(\textbf{h}^{H}_{d,k}+\textbf{f}^H_{k}\boldsymbol{\Psi}_{t}\textbf{G})\textbf{W}\textbf{s}+\textbf{f}^H_{k}\boldsymbol{\Psi}_{t}\textbf{v}+n_{k}, \enspace k\in\mathcal K_{t},
	\end{split}
\end{equation}

\noindent and the correspoding SINR can be given by:
\begin{equation}
	\label{eqn_5}
	\text{SINR}_{k}=\frac{|\tilde{\textbf{h}}^{H}_{t,k}\textbf{w}_{k}|^2}{\sum\limits_{i=1,i\neq k}\limits^{K+M}|\tilde{\textbf{h}}^{H}_{t,k}\textbf{w}_{i}|^2\!+\!\sigma^2_{v}\|\textbf{f}^{H}_{k}\boldsymbol{\Psi}_{t}\|^2\!+\!\sigma^2_{k}}, \; k\!\in\!\mathcal K_{t},
\end{equation}

\noindent where $\tilde{\textbf{h}}_{t,k}^{H}\triangleq \textbf{h}^{H}_{d,k}+\textbf{f}^{H}_{k}\boldsymbol{\Psi}_{t}\textbf{G}\in\mathbb{C}^{1\times M}$
represents the equivalent cascaded channels from DFBS to the $k$-th $t$-user.

Hence, the sum rate can be calculated as:
\begin{equation}
	\label{eqn_6}
	R_{\text{sum}}=\sum\limits_{k=1}\limits^{K}\text{log}_{2}(1+\text{SINR}_{k}).
\end{equation}

From the perspective of radar sensing, to obtain an enhanced target sensing performance, one common approach is to maximize the signal power directed towards the target while minimizing it in other directions. On this basis, the assessment of sensing capabilities is usually conducted by using the SINR of target echo. In addition, considering that DFBS needs to perform both communication and sensing simultaneously, we assume that DFBS operate in FD mode, which will cause SI \cite{ref45}. Note that SI can be suppressed to an acceptable level by employing contemporary SI cancellation (SIC) techniques, but there still exists residual SI \cite{ref46, ref47}. Thus, due to the existing multiple targets and SI leakage, the received echo signal of the $q$-th target $\boldsymbol{y}_q \in \mathbb{C}^{M \times 1}$ at the DFBS can be represented as:
\begin{equation}\label{eqn_7}
	\boldsymbol{y}_q\!=\!b \textbf{h}_{d,q} \textbf{h}_{d,q}^H \textbf{W} \textbf{s}+b \sum_{i=1, i \neq q}^Q \textbf{h}_{d,i} \textbf{h}_{d,i}^H \textbf{W s}\!+\!\textbf{H}_{SI} \textbf{W} \textbf{s}\!+\!\textbf{z}_q,
\end{equation}
where the scalar $b$ represents the target radar cross section (RCS) with $\mathbb{E}\left\{|b|^2\right\}=\xi^2$, $\textbf{h}_{d, q} \in \mathbb{C}^{M \times 1}$ denotes the channel between the DFBS and the $q$-th target, and $\textbf{z}_q \sim \mathcal{C} V\left(\mathbf{0}, \sigma_z^2 \mathbf{I}_M\right)$ is AWGN for the $q$-th target. $\textbf{H}_{SI} \in \mathbb{C}^{M \times M}$ represents the SI channel between the DFBS transmitter and receiver, and $\textbf{H}_{SI} \textbf{W} \textbf{s}$ is the leaked transmission signal which is also known as SI signal. It is noted that the DFBS-target links are LOS and the angle of arrival/departure (AoA/AoD) of interest is known a priori.

Typically, in order to capture the desired reflected signal of the $q$-th target, we apply a radar receive filter $\textbf{u}_q \in \mathbb{C}^{M \times 1}$ to process the echo signal $\boldsymbol{y}_q$, and the radar output signal for the $q$-th target at the DFBS can be expressed as:
\begin{equation}\label{eqn_8}
		\textbf{u}_q^H \boldsymbol{y}_q\!=\!\textbf{u}_q^H b \textbf{H}_q \textbf{W} \textbf{s}\!+\!\textbf{u}_q^H b \sum_{i=1,i \neq q}^Q \textbf{H}_i \textbf{W} \textbf{s}\!+\!\textbf{u}_q^H \textbf{H}_{S I} \textbf{W} \textbf{s}\!+\!\textbf{u}_q^H \textbf{z}_q,
\end{equation}
where $\textbf{H}_q \triangleq \textbf{h}_{d, q} \textbf{h}_{d, q}^H \in \mathbb{C}^{M \times M}$ is the equivalent channel of the $q$-th echo signal. Thus, the radar output SINR of the $q$-th target can be caculated as \eqref{eqn_9}, which is presented at the top of the next page. 
\begin{figure*}[!t] 
	\centering
	\begin{equation}\label{eqn_9}
		\begin{aligned}
			\text{SINR}_{q} & =\frac{\xi^2\left|\textbf{u}_q^H \textbf{H}_q \textbf{W} \textbf{s}\right|^2}{\xi^2 \sum_{i=1, i \neq q}^Q\left|\textbf{u}_q^H \textbf{H}_i \textbf{W} \textbf{s}\right|^2+\left|\textbf{u}_q^H \textbf{H}_{S I} \textbf{W s}\right|^2+\left|\textbf{u}_q^H \textbf{z}_q\right|^2}\\
			& =\frac{\xi^2 \textbf{u}_q^H \textbf{H}_q \textbf{W} \textbf{W}^H \textbf{H}_q^H \textbf{u}_q}{\textbf{u}_q^H\left(\xi^2 \sum_{i=1, i \neq q}^Q \textbf{H}_i \textbf{W} \textbf{W}^H \textbf{H}_i^H+\textbf{H}_{S I} \textbf{W} \textbf{W}^H \textbf{H}_{S I}^H+\sigma_z^2 \mathbf{I}_M\right) \textbf{u}_q}.
		\end{aligned}
	\end{equation}\hrule
\end{figure*}
\subsection{Problem Formulation}
In this paper, through jointly designing the DFBS transmit beamforming $\textbf{W}$, the radar receive filter $\{\textbf{u}_q\}$, $\forall q$, and the active STAR-RIS reflection \& transmission beamforming matrices, i.e., $\boldsymbol{\Psi}_{r}$ and $\boldsymbol{\Psi}_{t}$, we aim to maximize the achievable sum-rate, as well as satisfying the worst radar SINR requirement $\Gamma_{q}$, active STAR-RIS hardware constraints, the given power budget $P_{\text{max}}^{\text{R}}$ at the active STAR-RIS and $P_{\text{max}}^{\text{B}}$ at the DFBS. The optimization problem is thus formulated as:
\begin{subequations} \label{eqn_10}
	\begin{align}
	\mathcal P_{0}&: \;
	\max\limits_{\{\textbf{w}_j\},\{\textbf{u}_q\},\boldsymbol{\Psi}_{r},\boldsymbol{\Psi}_{t}}\sum\limits_{k=1}\limits^{K}\text{log}_{2}(1+\text{SINR}_{k})
	\\
	&\text{s.t.} \hspace{0.2em} \mathrm{C}_{1}: \text{SINR}_q\geqslant \Gamma_{q}, \forall q,
	\\
	&\hspace{1.6em} \mathrm{C}_{2}: \sum\limits_{j=1}\limits^{K+M}\|\textbf{w}_{j}\|^2\leqslant P_{\text{max}}^{\text{B}},
	\\
	\begin{split}
	&\hspace{1.8em} \mathrm{C}_{3}:
	\sum\limits_{j=1}\limits^{K+M}\|\boldsymbol{\Psi}_{r}\textbf{G}\textbf{w}_{j}\|^2+\sigma^2_{v}\|\boldsymbol{\Psi}_{r}\|^{2}_{F}+
	\\
	&\hspace{2em} \sum\limits_{j=1}\limits^{K+M}\|\boldsymbol{\Psi}_{t}\textbf{G}\textbf{w}_{j}\|^2+\sigma^2_{v}\|\boldsymbol{\Psi}_{t}\|^{2}_{F}
	\leqslant P_{\text{max}}^{\text{R}},
	\end{split}
    \\
    &\hspace{2.2em} \mathrm{C}_{4}:
    a_{r,n}\geqslant 0,\; a_{t,n}\geqslant 0, \; \forall n,
    \\
    &\hspace{2.3em} \mathrm{C}_{5}:
    |\phi_{r,n}|=1, \; |\phi_{t,n}|=1,\; \forall n.
	\end{align}
\end{subequations}
where, constraints $\mathrm{C}_{1}$ indicates the radar sensing SINR requirement per target, $\mathrm{C}_{2}$ and $\mathrm{C}_{3}$ represent the given maximum power budget at the DFBS and active STAR-RIS, respectively, $\mathrm{C}_{4}$ means the active nature limitation of STAR-RIS amplitude, and $\mathrm{C}_{5}$ indicates the unit-modulus constraint of phase shift. It is clear that $\mathcal P_{0}$ is a non-convex optimization problem due to the complex objective function (10a) which includes logarithmic and fractional terms, the interdependence of variables in both the objective function (10a) and the power constraint $\mathrm{C}_{3}$, and the constraint $\mathrm{C}_{5}$ which imposes a unit modulus condition. Next, we propose an efficient iterative algorithm to deal with it.

\section{Joint Beamforming Optimization }
In this section, an AO algorithm is developed to solve $\mathcal P_{0}$. Specifically, FP method is first utilized to convert the objective function into a more tractable form. After that, the transformed problem is divided into multiple sub-problems, that are solved in an alternating way.

\subsection{FP-based Transformation}
We first convert the complicated sum-of-logarithms and fractions in (10a) into an easier-to-compute feasible form by using the closed-form FP algorithm \cite{ref48,ref49}. Subsequently, by introducing an auxiliary variable $\boldsymbol{\gamma}=[\gamma_{1},\dots,\gamma_{K}]^{T}$ and utilizing the Lagrangian dual transform to reformulate the original objective function (10a) as:
\begin{equation}
	\label{eqn_11}
	\begin{split}
	\tilde{R}_{\text{sum}}(\boldsymbol{\gamma},\boldsymbol{\Psi}_{r},\boldsymbol{\Psi}_{t},\{\textbf{w}_j\})=\sum\limits_{k=1}\limits^{K+M}(1+\gamma_{k})-\sum\limits_{k=1}\limits^{K+M}\gamma_{k}
	\\
	\hspace{2.2em}+\sum\limits_{k\in\mathcal K_{r}}\frac{(1+\gamma_{k})|\tilde{\textbf{h}}_{r,k}^{H}\textbf{w}_{k}|^2}{\sum\limits_{j=1}\limits^{K+M}|\tilde{\textbf{h}}^{H}_{r,k}\textbf{w}_{j}|^2+\sigma^2_{v}\|\textbf{f}^{H}_{k}\boldsymbol{\Psi}_{r}\|^2+\sigma^2_{k}}
	\\
	\hspace{2.2em}+\sum\limits_{k\in\mathcal K_{t}}\frac{(1+\gamma_{k})|\tilde{\textbf{h}}_{t,k}^{H}\textbf{w}_{k}|^2}{\sum\limits_{j=1}\limits^{K+M}|\tilde{\textbf{h}}^{H}_{t,k}\textbf{w}_{j}|^2+\sigma^2_{v}\|\textbf{f}^{H}_{k}\boldsymbol{\Psi}_{t}\|^2+\sigma^2_{k}}.
	\end{split}
\end{equation}

Next, the quadratic transform is employed on the final two fractional terms of \eqref{eqn_11} through the introduction of another auxiliary variable $\boldsymbol{\rho}=[\rho_{1},\dots,\rho_{K}]^{T}$ to further recast them to \eqref{eqn_12}, which is presented at the top of the next page.
\begin{figure*}[!t] 
	\centering
\begin{equation}
	\label{eqn_12}
	\begin{split}
		f(\boldsymbol{\gamma},\boldsymbol{\Psi}_{r},\boldsymbol{\Psi}_{t},\textbf{w},\boldsymbol{\rho})&=\sum\limits_{k\in\mathcal K_{r}}\left(2\sqrt{\!1+\!\gamma_{k}}\Re\{\rho^{*}_{k}\tilde{\textbf{h}}_{r,k}^{H}\textbf{w}_{k}\}\!-\!|\rho_{k}|^2\left(\sum\limits_{j=1}\limits^{K+M}|\tilde{\textbf{h}}_{r,k}^{H}\textbf{w}_{j}|^2\!+\!\sigma^2_{v}\|\textbf{f}^{H}_{k}\boldsymbol{\Psi}_{r}\|^2\!+\!\sigma^2_{k}\right)\right)
	\\
		&+\sum\limits_{k\in\mathcal K_{t}}\left(2\sqrt{\!1+\!\gamma_{k}}\Re\{\rho^{*}_{k}\tilde{\textbf{h}}_{t,k}^{H}\textbf{w}_{k}\}\!-\!|\rho_{k}|^2\left(\sum\limits_{j=1}\limits^{K+M}|\tilde{\textbf{h}}_{t,k}^{H}\textbf{w}_{j}|^2\!+\!\sigma^2_{v}\|\textbf{f}^{H}_{k}\boldsymbol{\Psi}_{t}\|^2\!+\!\sigma^2_{k}\right)\right),
	\end{split}
\end{equation}\hrule
\end{figure*}

By substituting \eqref{eqn_12} into \eqref{eqn_11}, the original optimization problem $\mathcal P_{0}$ can be equivalently rewritten as $\mathcal P_{1}$:
\begin{subequations} \label{eqn_13}
	\begin{align}
		\mathcal P_{1}: \;
		&\max\limits_{\{\textbf{w}_j\},\{\textbf{u}_q\},\boldsymbol{\Psi}_{r},\boldsymbol{\Psi}_{t},\boldsymbol{\gamma},\boldsymbol{\rho}}\tilde{R}_{\text{sum}}(\boldsymbol{\gamma},\boldsymbol{\Psi}_{r},\boldsymbol{\Psi}_{t},\textbf{w},\boldsymbol{\rho})
		\\
		&\hspace{2em} \text{s.t.} \hspace{0.6em} \mathrm{C}_{1}, \mathrm{C}_{2}, \mathrm{C}_{3},\mathrm{C}_{4}, \mathrm{C}_{5}.
	\end{align}
\end{subequations}

Then, $\mathcal P_{1}$ can be addressed by adopting the AO framework, which involves dividing it into several sub-problems. 

\subsection{Optimize Auxiliary Variable Vectors $\boldsymbol{\gamma}$ and $\boldsymbol{\rho}$}

With given $\{\textbf{w}_j\}$, $\boldsymbol{\Psi}_{r}$ and $\boldsymbol{\Psi}_{t}$, the optimal $\boldsymbol{\gamma}$ and $\boldsymbol{\rho}$ can be determined explicitly by letting $\frac{\partial \tilde{R}_{\text{sum}}}{\partial \gamma_{k}}$ and
$\frac{\partial f}{\partial \rho_{k}}$ to zero, i.e.:
\begin{equation}
	\label{eqn_14}
	{\gamma_{k}^{opt}} = \begin{cases}
		\Large{\frac{|\tilde{\textbf{h}}^{H}_{r,k}\textbf{w}_{k}|^2}{\sum\limits_{i=1,i\neq k}\limits^{K+M}|\tilde{\textbf{h}}^{H}_{r,k}\textbf{w}_{i}|^2+\sigma^2_{v}\|\textbf{f}^{H}_{k}\boldsymbol{\Psi}_{r}\|^2+\sigma^2_{k}},}&{k\in\mathcal K_{r}},\\ 
		\Large{\frac{|\tilde{\textbf{h}}^{H}_{t,k}\textbf{w}_{k}|^2}{\sum\limits_{i=1,i\neq k}\limits^{K+M}|\tilde{\textbf{h}}^{H}_{t,k}\textbf{w}_{i}|^2+\sigma^2_{v}\|\textbf{f}^{H}_{k}\boldsymbol{\Psi}_{t}\|^2+\sigma^2_{k}},}&{k\in\mathcal K_{t}}.
	\end{cases}
\end{equation}
\begin{equation}
	\label{eqn_15}
	{\rho_{k}^{opt}} = \begin{cases}
		\Large{\frac{\sqrt{1+\gamma_{k}}\tilde{\textbf{h}}^{H}_{r,k}\textbf{w}_{k}}{\sum\limits_{j=1}\limits^{K+M}|\tilde{\textbf{h}}^{H}_{r,k}\textbf{w}_{j}|^2+\sigma^2_{v}\|\textbf{f}^{H}_{k}\boldsymbol{\Psi}_{r}\|^2+\sigma^2_{k}},}&{k\in\mathcal K_{r}},\\ 
		\Large{\frac{\sqrt{1+\gamma_{k}}\tilde{\textbf{h}}^{H}_{t,k}\textbf{w}_{k}}{\sum\limits_{j=1}\limits^{K+M}|\tilde{\textbf{h}}^{H}_{t,k}\textbf{w}_{j}|^2+\sigma^2_{v}\|\textbf{f}^{H}_{k}\boldsymbol{\Psi}_{t}\|^2+\sigma^2_{k}},}&{k\in\mathcal K_{t}}.
	\end{cases}
\end{equation}

\subsection{Optimize Radar Receive Beamforming $\{\textbf{u}_q\}$}

For given $\{\textbf{w}_j\}$, the radar receive beamforming optimization problem can be reduced to the typical generalized ``Rayleigh quotient" optimization problem \cite{ref33} as follows:
\begin{equation} 
	\label{eqn_16}
		\max\limits_{\textbf{u}_q} \quad \frac{\xi^2\textbf{u}_q^{H}\textbf{C}_q\textbf{u}}{\textbf{u}_q^{H}\textbf{E}_q\textbf{u}_q},
\end{equation}
whose optimal solution $\textbf{u}_q^{opt}$ is the eigenvector that corresponds to the maximum eigenvalue of matrix $\textbf{E}_q^{-1}\textbf{C}_q$, where $\textbf{C}_q=\textbf{H}_{q}\textbf{W}\textbf{W}^{H}\textbf{H}_{q}^{H}$, $\textbf{E}_q=\xi^2 \sum_{i=1, i \neq q}^Q \textbf{H}_i \textbf{W} \textbf{W}^H \textbf{H}_i^H+\textbf{H}_{S I} \textbf{W} \textbf{W}^H \textbf{H}_{S I}^H+\sigma_z^2 \mathbf{I}_M$.

\subsection{Optimize DFBS Beamforming $\textbf{W}$}

For given $\{\textbf{u}_q\}$, $\boldsymbol{\Psi}_{r}$, $\boldsymbol{\Psi}_{t}$, $\boldsymbol{\gamma}$ and $\boldsymbol{\rho}$, the optimization of DFBS transmit beamforming $\textbf{W}$ can be modeled as:

\begin{subequations} \label{eqn_17}
	\begin{align}
		&\max\limits_{\{\textbf{w}_j\}}\hspace{0.5em}\tilde{R}_{\text{sum}}(\boldsymbol{\gamma},\boldsymbol{\Psi}_{r},\boldsymbol{\Psi}_{t},\{\textbf{w}_j\},\boldsymbol{\rho})
		\\
		&\hspace{0.5em}\text{s.t.} \hspace{0.6em} \mathrm{C}_{1}, \mathrm{C}_{2}, \mathrm{C}_{3}.
	\end{align}
\end{subequations}

For convenience, we first give the following definitions: 
\begin{equation} 
	\label{eqn_18}
	\tilde{\textbf{w}}\triangleq\text{vec}\{\textbf{W}\},
\end{equation}
\begin{equation} 
	\label{eqn_19}
	\boldsymbol{\alpha}_{k}\triangleq\sqrt{1+\gamma_{k}}\rho_{k}\tilde{\textbf{h}}_{r/t,k}, \; \forall k\in\mathcal K,
\end{equation}
\begin{equation} 
	\label{eqn_20}
	\begin{split}
		\boldsymbol{\alpha}_{r}\triangleq[\boldsymbol{\alpha}_{1}^{T},\dots,\boldsymbol{\alpha}_{k}^{T},\textbf{0}_{1\times [M(K\!+\!M)\!-\!MK_{r}]}]^{T}, \forall k \in \mathcal K_{r},
	\end{split}
\end{equation}
\begin{equation} 
	\label{eqn_21}
		\textbf{Q}_{r}\triangleq\textbf{I}_{K+M}\otimes\sum\limits_{k=1}\limits^{K_{r}}|\rho_{k}|^2\tilde{\textbf{h}}_{r,k}\tilde{\textbf{h}}_{r,k}^{H},
\end{equation}
\begin{equation} 
	\label{eqn_22}
	\begin{split}
		\boldsymbol{\alpha}_{t}\triangleq[\textbf{0}_{1\times MK_{r}},\boldsymbol{\alpha}_{1}^{T},\dots,\boldsymbol{\alpha}_{k}^{T},\textbf{0}_{1\times MM]}]^{T},  \forall k \in \mathcal K_{t},
	\end{split}
\end{equation}
\begin{equation} 
	\label{eqn_23}
	\textbf{Q}_{t}\triangleq\textbf{I}_{K+M}\otimes\sum\limits_{k=1}\limits^{K_{t}}|\rho_{k}|^2\tilde{\textbf{h}}_{t,k}\tilde{\textbf{h}}_{t,k}^{H},
\end{equation}
\begin{equation} 
	\label{eqn_24}
	\textbf{Y}_q\triangleq\textbf{I}_{K+M}\otimes\textbf{H}_{q}^{H}\textbf{u}_t\textbf{u}_q^{H}\textbf{H}_{q},\forall q,
\end{equation}
\begin{equation} 
	\label{eqn_25}
	\textbf{Y}_{SI,q}\triangleq\textbf{I}_{K+M}\otimes\textbf{H}_{SI}^{H}\textbf{u}_q\textbf{u}_q^{H}\textbf{H}_{SI},\forall q,
\end{equation}
\begin{equation} 
	\label{eqn_26}
	\begin{split}
		\boldsymbol{\Xi}_{r}\triangleq(\textbf{I}_{K+M}\otimes\textbf{G}^{H}\boldsymbol{\Psi}_{r}^{H}\boldsymbol{\Psi}_{r}\textbf{G}),
		\\
		\boldsymbol{\Xi}_{t}\triangleq(\textbf{I}_{K+M}\otimes\textbf{G}^{H}\boldsymbol{\Psi}_{t}^{H}\boldsymbol{\Psi}_{t}\textbf{G}).
	\end{split}
\end{equation}

Then, we can reformulate the subproblem into explicit and compact forms with respect to $\tilde{\textbf{w}}$:
\begin{subequations} \label{eqn_27}
	\begin{align}
		\mathcal P_{2}:
		&\max\limits_{\tilde{\textbf{w}}}\hspace{0.3em}\Re\{2\boldsymbol{\alpha}_{r}^{H}\tilde{\textbf{w}}\}-\tilde{\textbf{w}}^{H}\textbf{Q}_{r}\tilde{\textbf{w}}+\Re\{2\boldsymbol{\alpha}_{t}^{H}\tilde{\textbf{w}}\}-\tilde{\textbf{w}}^{H}\textbf{Q}_{t}\tilde{\textbf{w}}
		\\
		\begin{split}
		\text{s.t.} & \mathrm{C}_{1}: \xi^2 \tilde{\textbf{w}}^H \textbf{Y}_q \tilde{\textbf{w}}\!-\!\xi^2 \Gamma_q \tilde{\textbf{w}}^H \!\sum_{i=1,i \neq q}^Q \! \textbf{Y}_i \tilde{\textbf{w}}\!-\!\Gamma_q \tilde{\textbf{w}}^H \textbf{Y}_{SI,q} \tilde{\textbf{w}} \\
		&\hspace{2em} \geq \Gamma_q \textbf{u}_q^H \sigma_z^2 \mathbf{I}_M \textbf{u}_q, \forall q,
	    \end{split}
		\\
		&\hspace{0.2em} \mathrm{C}_{2}: \|\tilde{\textbf{w}}\|^2 \leqslant P_{\text{max}}^{\text{B}},
		\\
		\begin{split}
		&\hspace{0.2em} \mathrm{C}_{3}:\tilde{\textbf{w}}^{H}\boldsymbol{\Xi}_{r}\tilde{\textbf{w}}\!+\!
		\tilde{\textbf{w}}^{H}\boldsymbol{\Xi}_{t}\tilde{\textbf{w}} \!\leqslant\! P_{\text{max}}^{\text{R}}\!-\!\sigma_{v}^2\|\boldsymbol{\Psi}_{r}\|^2_{F}\!-\!\sigma_{v}^2\|\boldsymbol{\Psi}_{t}\|^2_{F}.		
	\end{split}
	\end{align}
\end{subequations}

For simplicity to calculate, by mathematical transformations, $\mathrm{C}_{1}$ can be rewritten as $\tilde{\textbf{w}}^H \tilde{\textbf{Y}}_q \tilde{\textbf{w}} \geq \eta_q, \forall q$, where $\tilde{\textbf{Y}}_q=\frac{\textbf{Y}_q}{\Gamma_q}-\sum_{i=1, i \neq q}^Q \textbf{Y}_i-\frac{\textbf{Y}_{SI,q}}{\xi^2}$, $\eta_q=\textbf{u}_q^H \sigma_z^2 \mathbf{I}_M \textbf{u}_q/{\xi^2}$. Since $\tilde{\textbf{Y}}_q$ is a positive semi-definite Hermitian matrix, $\mathrm{C}_{1}$ is a non-convex constraint, which makes $\mathcal P_{2}$ non-convex. In order to deal with this formidable undertaking, we exploit the ﬁrst-order Taylor expansion \cite{ref50} to obtain an suitable surrogate function for $\tilde{\textbf{w}}^{H}\tilde{\textbf{Y}}_q\tilde{\textbf{w}}$ and transform $\mathrm{C}_{1}$ into a linear constraint, which is given by:
\begin{equation} 
	\label{eqn_28}
	\tilde{\textbf{w}}^{H}\tilde{\textbf{Y}}_q\tilde{\textbf{w}}\geqslant \tilde{\textbf{w}}^{H}_{s}\tilde{\textbf{Y}}_q\tilde{\textbf{w}}_{s}+2\Re\{\tilde{\textbf{w}}^{H}_{s}\tilde{\textbf{Y}}_q(\tilde{\textbf{w}}-\tilde{\textbf{w}}_{s})\}.
\end{equation}

Then, $\mathcal P_{2}$ can be rewritten as:
\begin{subequations} \label{eqn_29}
	\begin{align}
		\mathcal P_{3}&:
		\min\limits_{\tilde{\textbf{w}}}\hspace{0.3em}\tilde{\textbf{w}}^{H}\textbf{Q}_{r}\tilde{\textbf{w}}-\Re\{2\boldsymbol{\alpha}_{r}^{H}\tilde{\textbf{w}}\}+\tilde{\textbf{w}}^{H}\textbf{Q}_{t}\tilde{\textbf{w}}-\Re\{2\boldsymbol{\alpha}_{t}^{H}\tilde{\textbf{w}}\}
		\\
		&\text{s.t.} \hspace{0.2em} \mathrm{C}_{1}: \tilde{\textbf{w}}^{H}_{s}\tilde{\textbf{Y}}_q\tilde{\textbf{w}}_{s}+2\Re\{\tilde{\textbf{w}}^{H}_{s}\tilde{\textbf{Y}}_q(\tilde{\textbf{w}}-\tilde{\textbf{w}}_{s})\}\geqslant \eta_q, \forall q,
		\\
		&\hspace{1.5em} \mathrm{C}_{2},\mathrm{C}_{3}.
	\end{align}
\end{subequations}

One can observe that $\mathcal P_{3}$ is a standard QCQP problem, which can be effectively tackled with standard convex optimization algorithm or tools, e.g., CVX \cite{ref51}.

\subsection{Optimize Active STAR-RIS Beamforming $\boldsymbol{\Psi}_{r}$ and $\boldsymbol{\Psi}_{t}$}

In this section, three work modes are considered for the active STAR-RIS, and the active STAR-RIS beamforming is solved under each mode. 

1) UED mode

In UED mode, each element of the active STAR-RIS can transmit and reflect the incident signals with different amplitudes and phases simultaneously, and this mode owns the highest DoFs \cite{ref26}. With given other variables, considering the fact that $\tilde{\textbf{h}}_{r,k}^{H}\triangleq\textbf{h}_{d,k}^{H}+\textbf{f}^{H}_{k}\boldsymbol{\Psi}_{r}\textbf{G}=\textbf{h}_{d,k}^{H}+\boldsymbol{\psi}^{H}_{r}$diag$(\textbf{f}_{k})\textbf{G}$, $\forall k \in \mathcal K_{r}$ and $\tilde{\textbf{h}}_{t,k}^{H}\triangleq\textbf{h}_{d,k}^{H}+\textbf{f}^{H}_{k}\boldsymbol{\Psi}_{t}\textbf{G}=\textbf{h}_{d,k}^{H}+\boldsymbol{\psi}^{H}_{t}$diag$(\textbf{f}_{k})\textbf{G}$, $\forall k \in \mathcal K_{t}$, by defining \eqref{eqn_30}, \eqref{eqn_31}, \eqref{eqn_32}, \eqref{eqn_33} and \eqref{eqn_34}, as shown at the top of the next page,
this subproblem with respect to $\boldsymbol{\psi}_{r}$ and $\boldsymbol{\psi}_{t}$ can be reformulated as:  
\begin{figure*}[!t] 
	\centering
\begin{equation} 
	\label{eqn_30}
		\textbf{d}_{r} \triangleq\sum\limits_{k\in\mathcal K_{r}}\sqrt{1+\gamma_{k}}\text{diag}(\rho^{*}_{k}\textbf{f}^{H}_{k})\textbf{G}\textbf{w}_{k}-|\rho_{k}|^2\text{diag}(\textbf{f}_{k}^{H})\textbf{G}\sum\limits_{j=1}\limits^{K+M}\textbf{w}_{j}\textbf{w}^{H}_{j}\textbf{h}_{d,k},
\end{equation}\vspace{-0.66cm}
\end{figure*}
\begin{figure*}[!t] 
	\centering
	\begin{equation} 
		\label{eqn_31}
		\textbf{D}_{r} \triangleq\sum\limits_{k\in\mathcal K_{r}}|\rho_{k}|^2\sigma_{v}^2\text{diag}(\textbf{f}_{k}^{H})\text{diag}(\textbf{f}_{k})+|\rho_{k}|^2\sum\limits_{j=1}\limits^{K+M}\text{diag}(\textbf{f}_{k}^{H})\textbf{G}\textbf{w}_{j}\textbf{w}^{H}_{j}\textbf{G}^{H}\text{diag}(\textbf{f}_{k}),
	\end{equation}\vspace{-0.66cm}
\end{figure*}
\begin{figure*}[!t] 
	\centering
	\begin{equation} 
		\label{eqn_32}
			\textbf{d}_{t}\triangleq\sum\limits_{k\in\mathcal K_{t}}\sqrt{1+\gamma_{k}}\text{diag}(\rho^{*}_{k}\textbf{f}^{H}_{k})\textbf{G}\textbf{w}_{k}-|\rho_{k}|^2\text{diag}(\textbf{f}_{k}^{H})\textbf{G}\sum\limits_{j=1}\limits^{K+M}\textbf{w}_{j}\textbf{w}^{H}_{j}\textbf{h}_{d,k},
	\end{equation}\vspace{-0.66cm}
\end{figure*}
\begin{figure*}[!t] 
	\centering
	\begin{equation} 
		\label{eqn_33}
		\textbf{D}_{t}\triangleq\sum\limits_{k\in\mathcal K_{t}}|\rho_{k}|^2\sigma_{v}^2\text{diag}(\textbf{f}_{k}^{H})\text{diag}(\textbf{f}_{k})+|\rho_{k}|^2\sum\limits_{j=1}\limits^{K+M}\text{diag}(\textbf{f}_{k}^{H})\textbf{G}\textbf{w}_{j}\textbf{w}^{H}_{j}\textbf{G}^{H}\text{diag}(\textbf{f}_{k}),
	\end{equation}\vspace{-0.66cm}
\end{figure*}
\begin{figure*}[!t] 
	\centering
	\begin{equation} 
		\label{eqn_34}
		\boldsymbol{\Pi}=\sum\limits_{j=1}\limits^{K+M}\text{diag}(\textbf{G}\textbf{w}_{j})(\text{diag}(\textbf{G}\textbf{w}_{j}))^{H}+\sigma_{v}^2\textbf{I}_{N}.
	\end{equation}\hrule
\end{figure*}
\begin{subequations} \label{eqn_35}
	\begin{align}
		\mathcal P_{4}:
		&\min\limits_{\boldsymbol{\psi}_{r},\boldsymbol{\psi}_{t}}
		\boldsymbol{\psi}_{r}^{H}\textbf{D}_{r}\boldsymbol{\psi}_{r}\!-\!\Re\{2\boldsymbol{\psi}_{r}^{H}\textbf{d}_{r}\}+\boldsymbol{\psi}_{t}^{H}\textbf{D}_{t}\boldsymbol{\psi}_{t}\!-\!\Re\{2\boldsymbol{\psi}_{t}^{H}\textbf{d}_{t}\}
		\\
		&\hspace{0.5em} \text{s.t.} \hspace{0.6em} \mathrm{C}_{3}: \boldsymbol{\psi}_{r}^{H}\boldsymbol{\Pi}\boldsymbol{\psi}_{r}+\boldsymbol{\psi}_{t}^{H}\boldsymbol{\Pi}\boldsymbol{\psi}_{t}\leqslant P_{\text{max}}^{\text{R}}.
	\end{align}
\end{subequations}

Obviously, $\mathcal P_{4}$ is also a QCQP problem that can be addressed with the standard convex optimization algorithm. By considering the constraints $\mathrm{C}_{4}$ and $\mathrm{C}_{5}$, the associated phase-shift matrix $\boldsymbol{\Phi}_{r}^{opt}\in\mathbb{C}^{N\times N}$, $\boldsymbol{\Phi}_{t}^{opt}\in\mathbb{C}^{N\times N}$ and ampliﬁcation factor vector $\boldsymbol{a}_{r}^{opt}\in\mathbb{C}^{N\times 1}$, $\boldsymbol{a}_{t}^{opt}\in\mathbb{C}^{N\times 1}$ are given by:
 \begin{subequations} \label{eqn_36}
 	\begin{align}
 		\boldsymbol{\Phi}_{r}^{opt} &= \text{diag}(\text{exp}(j\text{arg}(\boldsymbol{\psi}_{r}^{opt}))),
 		\\
 		\boldsymbol{\Phi}_{t}^{opt} &= \text{diag}(\text{exp}(j\text{arg}(\boldsymbol{\psi}_{t}^{opt}))),
 		\\
 		\boldsymbol{a}_{r}^{opt} &= \text{diag}(\text{exp}(-j\text{arg}(\boldsymbol{\psi}_{r}^{opt})))\boldsymbol{\psi}_{r}^{opt},
 		\\
 		\boldsymbol{a}_{t}^{opt} &= \text{diag}(\text{exp}(-j\text{arg}(\boldsymbol{\psi}_{t}^{opt})))\boldsymbol{\psi}_{t}^{opt}.
 	\end{align}
 \end{subequations} 

According to the definition of the reflection and transmission precoding matrices for active STAR-RIS in Part II.A, i.e., $\boldsymbol{\Psi}_{r}=\text{diag}(\boldsymbol{\psi}_{r})$ and $\boldsymbol{\Psi}_{t}=\text{diag}(\boldsymbol{\psi}_{t})$, we can obtain the optimal $\boldsymbol{\Psi}_{r}$ and $\boldsymbol{\Psi}_{t}$, respectively.

We summarize the joint beamforming optimizing scheme for UED mode in Algorithm 1.   
\begin{algorithm}[H]
	\caption{Joint Beamforming Optimization for UED Mode.}\label{alg:alg1}
	\begin{algorithmic}
		\STATE 
		\STATE \textbf{Input}: $\textbf{h}_{d,k}$, $\textbf{f}_{k}$, $\sigma_{k}^2$, $\forall k\in\mathcal K$, $\textbf{G}$, $\textbf{h}_{d,q}$, $\sigma_{v}^2$, $\sigma_{z}^2$, $P_{\text{max}}^{\text{B}}$, $P_{\text{max}}^{\text{R}}$, $\Gamma_{t}$, $Q_{\text{max}}$, $\delta_{th}$  
		\STATE \textbf{Output}: DFBS transmit beamforming $\textbf{W}$, radar receive filter $\{\textbf{u}_q\}$, active STAR-RIS beamforming $\boldsymbol{\Psi}_{r}$ and $\boldsymbol{\Psi}_{t}$, and sum rate $R_{\text{sum}}$.
		\STATE 1: Initialize: $\textbf{W}$, $\boldsymbol{\Psi}_{r}$, $\boldsymbol{\Psi}_{t}$, $t=1$, $\delta=\infty$, $R_{\text{sum}}=0$
		\STATE 2: $\textbf{while}$ $t\leqslant Q_{\text{max}}$ and $\delta\geqslant\delta_{th}$ $\textbf{do}$
		\STATE 3: \hspace{0.2cm} $R_{\text{pre}}=R_{\text{sum}}$;     
		\STATE 4: \hspace{0.2cm} Update $\gamma_{k}^{opt}$ and $\rho_{k}^{opt}$, $\forall k\in\mathcal K$ by (14) and (15);
		\STATE 5: \hspace{0.2cm} Update radar receiver $\{\textbf{u}_q\}$ by solving (16);
		\STATE 6: \hspace{0.2cm} Update DFBS beamforming $\textbf{W}$ by solving $\mathcal P_{3}$;
		\STATE 7: \hspace{0.2cm} Update $\boldsymbol{\Psi}_{r}$ and $\boldsymbol{\Psi}_{t}$ by solving $\mathcal P_{4}$;
		\STATE 8: \hspace{0.2cm} Update $R_{\text{sum}}$ by (6);
		\STATE 9: \hspace{0.2cm} $\delta=\frac{|R_{\text{pre}}-R_{\text{sum}}|}{R_{\text{sum}}}$;
		\STATE 10: \hspace{0.1cm} $t=t+1$.
		\STATE 11: $\textbf{end while}$      		
	\end{algorithmic}
	\label{alg1}
\end{algorithm}

2) EED mode

In EED mode, the signals reflected and transmitted by active STAR-RIS have the same power, i.e., all elements of active STAR-RIS transmit and reflect the incident signal with the same amplitude simultaneously, given by 
$\textbf{A}_{r}=\textbf{A}_{t}=\textbf{A}\triangleq\text{diag}(\boldsymbol{a})=\text{diag}([a_{1},\dots,a_{N}]^{T})\in\mathbb{C}^{N \times N}$.

Recall that $\boldsymbol{\Psi}_{r}=\text{diag}(\boldsymbol{\psi}_{r})=\text{diag}(\boldsymbol{\Phi}_{r}\boldsymbol{a})$, $\boldsymbol{\Psi}_{t}=\text{diag}(\boldsymbol{\psi}_{t})=\text{diag}(\boldsymbol{\Phi}_{t}\boldsymbol{a})$, thus the equivalent channels $\tilde{\textbf{h}}_{r,k}^{H}$ and $\tilde{\textbf{h}}_{t,k}^{H}$ can be rephrased as $\tilde{\textbf{h}}_{r,k}^{H}\triangleq\textbf{h}_{d,k}^{H}+\textbf{f}_{k}^{H}\boldsymbol{\Psi}_{r}\textbf{G}=\textbf{h}_{d,k}^{H}+\boldsymbol{a}^{H}\text{diag}(\textbf{f}_{k}^{H}\boldsymbol{\Phi}_{r})\textbf{G}, \forall k \in \mathcal K_{r}$ and $\tilde{\textbf{h}}_{t,k}^{H}\triangleq\textbf{h}_{d,k}^{H}+\textbf{f}_{k}^{H}\boldsymbol{\Psi}_{t}\textbf{G}=\textbf{h}_{d,k}^{H}+\boldsymbol{a}^{H}\text{diag}(\textbf{f}_{k}^{H}\boldsymbol{\Phi}_{t})\textbf{G}, \forall k \in \mathcal K_{t}$. Let $\textbf{c}_{r}=\textbf{f}_{k}^{H}\boldsymbol{\Phi}_{r}\in\mathbb{C}^{1 \times N}, \forall k \in \mathcal K_{r}$ and $\textbf{c}_{t}=\textbf{f}_{k}^{H}\boldsymbol{\Phi}_{t}\in\mathbb{C}^{1 \times N}, \forall k \in \mathcal K_{t}$, we define \eqref{eqn_37}, \eqref{eqn_38}, \eqref{eqn_39} and \eqref{eqn_40}, as shown at the top of the next page.
\begin{figure*}[!t] 
	\centering
	\begin{equation} 
		\label{eqn_37}
		\boldsymbol{\mu}_{r}\triangleq\sum\limits_{k\in\mathcal K_{r}}\sqrt{1+\gamma_{k}}\text{diag}(\rho^{*}_{k}\textbf{c}_{r})\textbf{G}\textbf{w}_{k}-|\rho_{k}|^2\text{diag}(\textbf{c}_{r})\textbf{G}\sum\limits_{j=1}\limits^{K+M}\textbf{w}_{j}\textbf{w}^{H}_{j}\textbf{h}_{d,k},
	\end{equation}\vspace{-0.7cm}
\end{figure*}
\begin{figure*}[!t] 
	\centering
	\begin{equation} 
		\label{eqn_38}
		 \boldsymbol{\Omega}_{r} \triangleq\sum\limits_{k\in\mathcal K_{r}}|\rho_{k}|^2\sigma_{v}^2\text{diag}(\textbf{f}_{k}^{H})\text{diag}(\textbf{f}_{k})+|\rho_{k}|^2\sum\limits_{j=1}\limits^{K+M}\text{diag}(\textbf{c}_{r})\textbf{G}\textbf{w}_{j}\textbf{w}^{H}_{j}\textbf{G}^{H}\text{diag}(\textbf{c}_{r}^{H}),
	\end{equation}\vspace{-0.7cm}
\end{figure*}
\begin{figure*}[!t] 
	\centering
	\begin{equation} 
		\label{eqn_39}
		\boldsymbol{\mu}_{t}\triangleq\sum\limits_{k\in\mathcal K_{t}}\sqrt{1+\gamma_{k}}\text{diag}(\rho^{*}_{k}\textbf{c}_{t})\textbf{G}\textbf{w}_{k}-|\rho_{k}|^2\text{diag}(\textbf{c}_{t})\textbf{G}\sum\limits_{j=1}\limits^{K+M}\textbf{w}_{j}\textbf{w}^{H}_{j}\textbf{h}_{d,k},
	\end{equation}\vspace{-0.7cm}
\end{figure*}
\begin{figure*}[!t] 
	\centering
	\begin{equation} 
		\label{eqn_40}
		\boldsymbol{\Omega}_{t}\triangleq\sum\limits_{k\in\mathcal K_{t}}|\rho_{k}|^2\sigma_{v}^2\text{diag}(\textbf{f}_{k}^{H})\text{diag}(\textbf{f}_{k})+|\rho_{k}|^2\sum\limits_{j=1}\limits^{K+M}\text{diag}(\textbf{c}_{t})\textbf{G}\textbf{w}_{j}\textbf{w}^{H}_{j}\textbf{G}^{H}\text{diag}(\textbf{c}_{t}^{H}).
	\end{equation}\hrule
\end{figure*} 

With given other variables, the subproblem can be transformed into the following one:
\begin{subequations} \label{eqn_41}
	\begin{align}
		\mathcal P_{5}:
		&\min\limits_{\boldsymbol{a}}\boldsymbol{a}^{H}\boldsymbol{\Omega}_{r}\boldsymbol{a}-
		\Re\{2\boldsymbol{a}^{H}\boldsymbol{\mu}_{r}\}+
		\boldsymbol{a}^{H}\boldsymbol{\Omega}_{t}\boldsymbol{a}-
		\Re\{2\boldsymbol{a}^{H}\boldsymbol{\mu}_{t}\}
		\\
		&\hspace{0.5em} \text{s.t.} \hspace{0.6em} \mathrm{C}_{3}: \boldsymbol{a}^{H}\boldsymbol{\Pi}\boldsymbol{a}+\boldsymbol{a}^{H}\boldsymbol{\Pi}\boldsymbol{a}\leqslant P_{\text{max}}^{\text{R}},
		\\
		 &\hspace{2.3em} \mathrm{C}_{4}.
	\end{align}
\end{subequations}

It is clear that $\mathcal P_{5}$ is also a standard QCQP problem and we can easily obtain the optimal $\boldsymbol{a}$ by existing convex optimization algorithm or optimization tools.

After obtaining $\boldsymbol{a}$, to solve $\boldsymbol{\Psi}_{r}$ and $\boldsymbol{\Psi}_{t}$, considering that $\boldsymbol{\Psi}_{r}=\text{diag}(\boldsymbol{\psi}_{r})=\text{diag}(\textbf{A}\boldsymbol{\phi}_{r})$ and 
$\boldsymbol{\Psi}_{t}=\text{diag}(\boldsymbol{\psi}_{t})=\text{diag}(\textbf{A}\boldsymbol{\phi}_{t})$, we should optimize $\boldsymbol{\phi}_{r}$ and $\boldsymbol{\phi}_{t}$ next. Similarly, the equivalent channel $\tilde{\textbf{h}}_{r,k}^{H}$ and $\tilde{\textbf{h}}_{t,k}^{H}$ firstly be rewritten as 
$\tilde{\textbf{h}}_{r,k}^{H}\triangleq\textbf{h}_{d,k}^{H}+\textbf{f}_{k}^{H}\boldsymbol{\Psi}_{r}\textbf{G}=
\textbf{h}_{d,k}^{H}+\boldsymbol{\phi}_{r}^{H}\text{diag}(\textbf{f}_{k}^{H}\textbf{A})\textbf{G},\forall k \in \mathcal K_{r}$ and
$\tilde{\textbf{h}}_{t,k}^{H}\triangleq\textbf{h}_{d,k}^{H}+\textbf{f}_{k}^{H}\boldsymbol{\Psi}_{t}\textbf{G}=
\textbf{h}_{d,k}^{H}+\boldsymbol{\phi}_{t}^{H}\text{diag}(\textbf{f}_{k}^{H}\textbf{A})\textbf{G}, \forall k \in \mathcal K_{t}$. Let $\textbf{q}_{r}=\textbf{f}_{k}^{H}\textbf{A}\in\mathbb{C}^{1 \times N}, \forall k \in \mathcal K_{r}$ and $\textbf{q}_{t}=\textbf{f}_{k}^{H}\textbf{A}\in\mathbb{C}^{1 \times N}, \forall k \in \mathcal K_{t}$. Considering that the phase-shifts $\boldsymbol{\phi}_{r}$ and $\boldsymbol{\phi}_{t}$ are independent, we can optimize them separately. Next, we define:
\begin{equation} 
	\label{eqn_42}
	\begin{split}
		\hat{\boldsymbol{\mu}}_{r} &\triangleq\sum\limits_{k\in\mathcal K_{r}}\sqrt{1+\gamma_{k}}\text{diag}(\rho^{*}_{k}\textbf{q}_{r})\textbf{G}\textbf{w}_{k}
		\\
		&-|\rho_{k}|^2\text{diag}(\textbf{q}_{r})\textbf{G}\sum\limits_{j=1}\limits^{K+M}\textbf{w}_{j}\textbf{w}^{H}_{j}\textbf{h}_{d,k},
	\end{split}
\end{equation}
\begin{equation} 
	\label{eqn_43}
	\begin{split}
		\hspace{1cm} &\hat{\boldsymbol{\Omega}}_{r} \triangleq\sum\limits_{k\in\mathcal K_{r}}|\rho_{k}|^2\sigma_{v}^2\text{diag}(\textbf{f}_{k}^{H})\text{diag}(\textbf{f}_{k})+
		\\
		&|\rho_{k}|^2\sum\limits_{j=1}\limits^{K+M}\text{diag}(\textbf{q}_{r})\textbf{G}\textbf{w}_{j}\textbf{w}^{H}_{j}\textbf{G}^{H}\text{diag}(\textbf{q}_{r}^{H}).
	\end{split}
\end{equation}     

With given other variables, the optimization of $\boldsymbol{\phi}_{r}$ is equivalent to solve the problem that follows:
\begin{subequations} \label{eqn_44}
	\begin{align}
		\mathcal P_{6}: \;
		&\min\limits_{\boldsymbol{\phi}_{r}} \hspace{0.8em} \boldsymbol{\phi}^{H}_{r}\hat{\boldsymbol{\Omega}}_{r}\boldsymbol{\phi}_{r}+
		\Re\{2\boldsymbol{\phi}^{H}_{r}(-\hat{\boldsymbol{\mu}}_{r})\}
		\\
		&\hspace{0.5em} \text{s.t.} \hspace{0.6em} |\phi_{r,n}|=1, \; \forall n.
	\end{align}
\end{subequations}

The presence of the unit modulus constraint in (44b) results in $\mathcal P_{6}$ a non-convex optimization problem. Fortunately, two methods have been discussed in \cite{ref52} to solve this problem, i.e., MM and CCM. For a more comprehensive understanding, readers can consult \cite{ref52} for more information.\\
\indent By repeating the same procedure, $\boldsymbol{\phi}_{t}$ can also be optimized. Specifically, by defining:
\begin{equation} 
	\label{eqn_45}
	\begin{split}
		\hat{\boldsymbol{\mu}}_{t} &\triangleq\sum\limits_{k\in\mathcal K_{t}}\sqrt{1+\gamma_{k}}\text{diag}(\rho^{*}_{k}\textbf{q}_{t})\textbf{G}\textbf{w}_{k}
		\\
		&-|\rho_{k}|^2\text{diag}(\textbf{q}_{t})\textbf{G}\sum\limits_{j=1}\limits^{K+M}\textbf{w}_{j}\textbf{w}^{H}_{j}\textbf{h}_{d,k},
	\end{split}
\end{equation}
\begin{equation} 
	\label{eqn_46}
	\begin{split}
	\hspace{1cm} \hat{\boldsymbol{\Omega}}_{t} &\triangleq\sum\limits_{k\in\mathcal K_{t}}|\rho_{k}|^2\sigma_{v}^2\text{diag}(\textbf{f}_{k}^{H})\text{diag}(\textbf{f}_{k})+
		\\
		&|\rho_{k}|^2\sum\limits_{j=1}\limits^{K+M}\text{diag}(\textbf{q}_{t})\textbf{G}\textbf{w}_{j}\textbf{w}^{H}_{j}\textbf{G}^{H}\text{diag}(\textbf{q}_{t}^{H}),
	\end{split}
\end{equation} 

\noindent the corresponding optimization of $\boldsymbol{\phi}_{t}$ can be modeled as:
\begin{subequations} \label{eqn_47}
	\begin{align}
		\mathcal P_{7}: \;
		&\min\limits_{\boldsymbol{\phi}_{t}} \hspace{0.8em} \boldsymbol{\phi}^{H}_{t}\hat{\boldsymbol{\Omega}}_{t}\boldsymbol{\phi}_{t}+
		\Re\{2\boldsymbol{\phi}^{H}_{t}(-\hat{\boldsymbol{\mu}}_{t})\}
		\\
		&\hspace{0.5em} \text{s.t.} \hspace{0.6em} |\phi_{t,n}|=1, \; \forall n.
	\end{align}
\end{subequations}

We summarize the joint beamforming optimization scheme for UED mode in Algorithm 2.
\begin{algorithm}[ht]
	\caption{Joint Beamforming Optimization for EED Mode.}\label{alg:alg1}
	\begin{algorithmic}
		\STATE 
		\STATE \textbf{Input}: $\textbf{h}_{d,k}$, $\textbf{f}_{k}$, $\sigma_{k}^2$, $\forall k\in\mathcal K$, $\textbf{G}$, $\textbf{h}_{d,q}$, $\sigma_{v}^2$, $\sigma_{z}^2$, $P_{\text{max}}^{\text{B}}$, $P_{\text{max}}^{\text{R}}$, $\Gamma_{t}$, $Q_{\text{max}}$, $\delta_{th}$
		\STATE \textbf{Output}: DFBS transmit beamforming $\textbf{W}$, radar receive filter $\{\textbf{u}_q\}$, active STAR-RIS beamforming $\boldsymbol{\Psi}_{r}$ and $\boldsymbol{\Psi}_{t}$, and sum rate $R_{\text{sum}}$. 
		\STATE 1: Initialize: $\textbf{W}$, $\boldsymbol{\Psi}_{r}$, $\boldsymbol{\Psi}_{t}$, $t=1$, $\delta=\infty$, $R_{\text{sum}}=0$
		\STATE 2: $\textbf{while}$ $t\leqslant Q_{\text{max}}$ and $\delta\geqslant\delta_{th}$ $\textbf{do}$
		\STATE 3: \hspace{0.2cm} $R_{\text{pre}}=R_{\text{sum}}$;     
		\STATE 4: \hspace{0.2cm} Update $\gamma_{k}^{opt}$ and $\rho_{k}^{opt}$, $\forall k\in\mathcal K$ by (14) and (15);
		\STATE 5: \hspace{0.2cm} Update radar receiver $\{\textbf{u}_q\}$ by solving (16);
		\STATE 6: \hspace{0.2cm} Update DFBS beamforming $\textbf{W}$ by solving $\mathcal P_{3}$;
		\STATE 7: \hspace{0.2cm} Update $\boldsymbol{a}$ by solving $\mathcal P_{5}$
		\STATE 8: \hspace{0.2cm} Update $\boldsymbol{\Psi}_{r}$ and $\boldsymbol{\Psi}_{t}$ by solving $\mathcal P_{6}$ and $\mathcal P_{7}$;
		\STATE 9: \hspace{0.2cm} Calculate $\boldsymbol{\Phi}_{r}=\text{diag}(\boldsymbol{\Phi}_{r}\boldsymbol{a})$ and 
		$\boldsymbol{\Phi}_{t}=\text{diag}(\boldsymbol{\Phi}_{t}\boldsymbol{a})$;
		\STATE 10: \hspace{0.1cm} Update $R_{\text{sum}}$ by (6);
		\STATE 11: \hspace{0.1cm} $\delta=\frac{|R_{\text{pre}}-R_{\text{sum}}|}{R_{\text{sum}}}$;
		\STATE 12: \hspace{0.1cm} $t=t+1$.
		\STATE 13: $\textbf{end while}$      		
	\end{algorithmic}
	\label{alg1}
\end{algorithm}

3) SD mode

In SD mode, elements of active STAR-RIS are segregated into two distinct groups, with one group operating in reflection mode and the other group operating in transmission mode, i.e., the corresponding amplitude coefficient $a_{r,n}$ \& $a_{t,n}$ is fixed as 1 or 0. The algorithm for this mode can be obtained with the same algorithmic framework in EED mode as above with fixing the amplitude coefficient as 1 or 0. To make this paper more concise, we have omitted the details here.

\subsection{Analysis of Computational Complexity}

Based on the above proposed method, the joint DFBS beamforming and active STAR-RIS beamforming is straightforward. Next, we give a brief complexity analysis of the proposed algorithm. In each iteration, the computational complexity of updating $\boldsymbol{\gamma}$ is $\mathcal{O}(K^2N^2)$. The complexity of updating  $\boldsymbol{\rho}$ is $\mathcal{O}(K(K+1)N^2)$. The complexity of updating the receiver filter $\{\textbf{u}_q\}$ is $\mathcal{O}(QM^3)$. 
Given the complexity of finding a solution for the standard convex QCQP problem, the computational complexity of updating the transmit beamformer $\textbf{W}$ is $\mathcal{O}(\sqrt{MK+2}(1+MK)M^3K^3))$ and updating the active STAR-RIS precoding matrix $\boldsymbol{\Psi}_{r}$ and $\boldsymbol{\Psi}_{t}$ is $\mathcal{O}(2\sqrt{N+1}(1+2)N^3))$. Similarly, the computational complexity of updating $\textbf{A}$ is $\mathcal{O}(\sqrt{N+1}(1+2)N^3))$. For EED mode, the complexity of MM or CCM is given by $\mathcal{O}(N^3+I_{o}N^2)$, where $I_{o}$ represents the number of iterations necessary for the two algorithms to converge. Thus, the overall computational complexity of Algorithm 1 is approximated as the order of $\mathcal{O}(I_{c1}(K^2N^2+QM^3+M^{4.5}K^{4.5}+N^{4.5}+I_{o}N^2))$, where $I_{c1}$ represents the number of iterations necessary for Algorithm 1 to achieve convergence. Similarly, the overall computational complexity of Algorithm 2 is approximated as the order of $\mathcal{O}(I_{c2}(K^2N^2+QM^3+M^{4.5}K^{4.5}+N^{4.5}+I_{o}N^2))$, where $I_{c2}$ represents the number of iterations necessary for Algorithm 2 to achieve convergence.
\begin{figure}[!t]
	\centering
	\includegraphics[height=2.8in,width=3.4in]{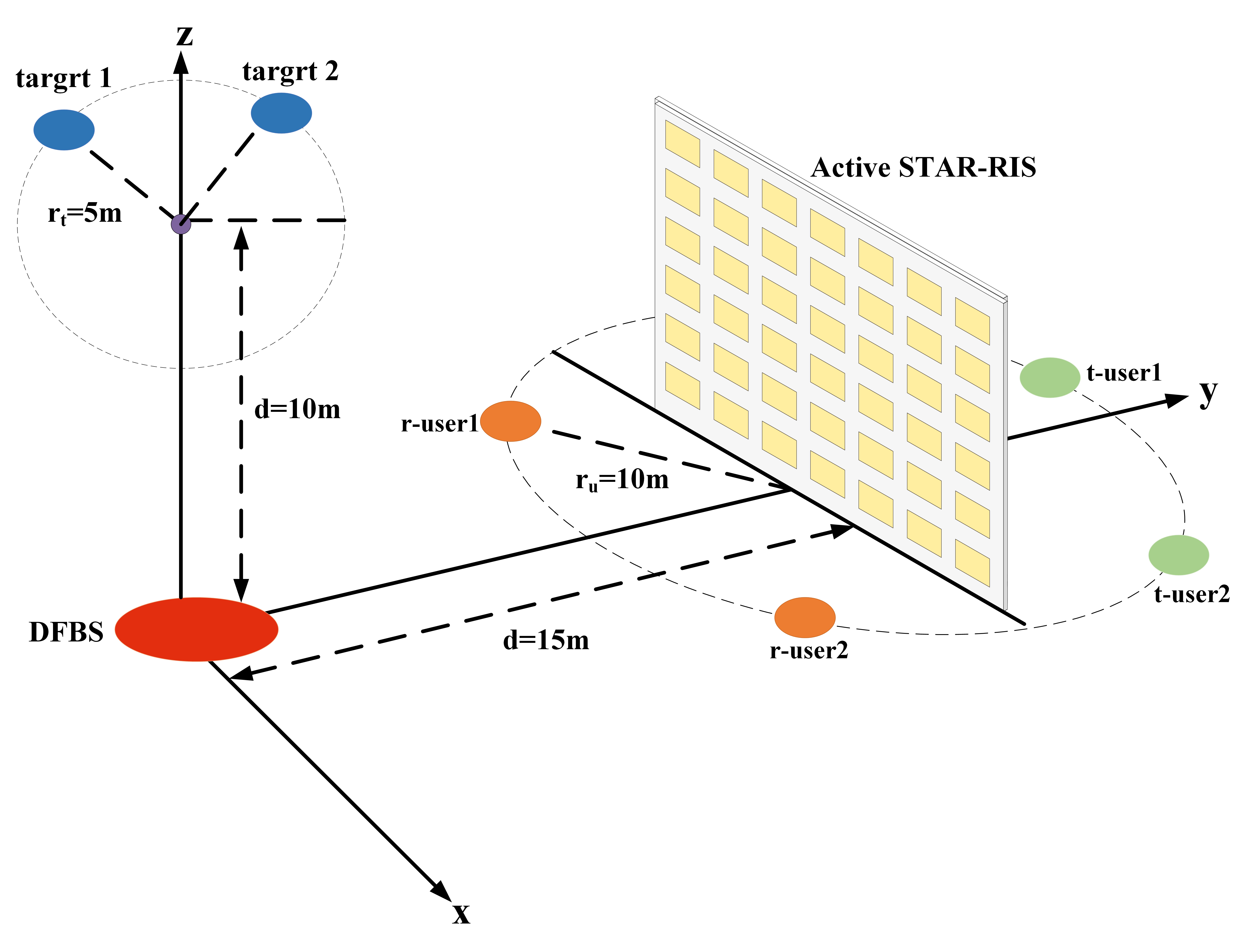}
	\caption{Simulation setup for active STAR-RIS-assisted ISAC system.}
	\label{fig_3}
\end{figure}
\section{Numerical Results}

In this section, simulation results are conducted to evaluate the performance of the proposed schemes. As depicted in Fig. \ref{fig_3}, in order to accurately depict the locations of DFBS and STAR-RIS, a three-dimensional coordinate system is utilized. Concretely, the DFBS is positioned at the origin (0, 0, 0) meters(m), while the active STAR-RIS is positioned at (0, 15, 0)m. On both sides of the active STAR-RIS, assuming two users are randomly distributed within a semicircular area with a radius of 10m centred around the active STAR-RIS, i.e., $K=4$. The two targets are assumed to be hovering within a circular area of a radius with 5m, centred around DFBS. We set the number of DFBS transmitting and receiving antennas to be the same, i.e., $M=8$. The number of active STAR-RIS elements is modeled as $N=64$, and the total transmit power budget is se to $P_{\text{max}}^{\text{total}}=16$dBm. For simplicity, we set the RCS as $\xi^2=1$, the sensing SINR constraint for all targets as $\Gamma_{t}=5$dB, and all noise power as the same, i.e., $\sigma_k^2=\sigma_v^2=\sigma_z^2=$ $-80$dBm, $\forall k\in\mathcal K$. Following \cite{ref32}, we employ the Rician channel model for all channels except the SI channel with $\textbf{H}=\text{PL}(\sqrt{\frac{\kappa}{\kappa+1}}\textbf{H}_{\text{LoS}}+\sqrt{\frac{1}{\kappa+1}}\textbf{H}_{\text{NLoS}})$, where $\text{PL}=\sqrt{L_0(\frac{d}{d_0})^{-\tau}}$, $\tau \in \{\tau_{BR}, \tau_{BQ}, \tau_{RU,k}, \tau_{BU,k}\}$, indicates the path loss associated with $\textbf{H}$, $d$ indicates the distance between two devices, $d_0=1$m is the path loss of reference distance, and $\kappa=1$ is the Rician factor. For the residual SI channel $\textbf{H}_{SI}$ at the DFBS, we follow \cite{ref45} to model it. The specific parameters setting can be found in TABLE II. Unless otherwise stated, the above experimental parameters remain unchanged throughout our simulations. 
\begin{table}[!t]
	\caption{System Parameters\label{tab:table2}}
	\centering
	\resizebox{0.5\textwidth}{!}{
		\begin{tabular}{|m{5.5cm}<{\centering}||m{4.8cm}<{\centering}|}
			\hline
			\textbf{Paramater} & \textbf{Value}\\
			\hline
			The number of antennas at DFBS & $M=8$\\
			\hline
			The number of active STAR-RIS elements & $N=64$\\
			\hline
			The number of communication users & $K=4$\\
			\hline
			The number of targets & $Q=2$\\
			\hline
			Path loss exponents & $\tau_{BQ}=\tau_{RU,k}=1.8$, $\tau_{BR}=3$, $\tau_{BU,k}=3.5$, $\forall k\in\mathcal K$\\
			\hline
			The path loss at 1 m & $L_0=30$dB\\
			\hline
			Rician factor & $\kappa=1$\\
			\hline
			The total transmit power budget & $P_{\text{max}}^{\text{total}}=16$dBm \\
			\hline
			The noise power & $\sigma^2_k=\sigma^2_v=\sigma^2_z=-80$dBm, $\forall k\in\mathcal K$\\
			\hline
			The target radar cross section & $\xi^2=1$  \\
			\hline
			The sensing SINR requirement & $\Gamma_{t}=5$dB \\
			\hline
		\end{tabular}
	}
\end{table}

To verify the effectiveness of our proposed schemes, we conduct a comparison with the following baseline:

\begin{itemize}
	\item{\textbf{The passive STAR-RIS-assisted ISAC system scheme (legend ``passive STAR-RIS ES'')}: As verified by \cite{ref28,ref29}, the optimal mode for passive STAR-RIS is energy splitting (ES), which means that the transmission and reflection coefficients of each STAR-RIS element can be optimized, and a high degree of flexibility for system design is enabled. It is the same as the UED mode in active STAR-RIS. Therefore, we adopt the ES mode for passive STAR-RIS.}
\end{itemize}
\begin{figure}[!t]
	\centering
	\includegraphics[height=2.8in,width=3.4in]{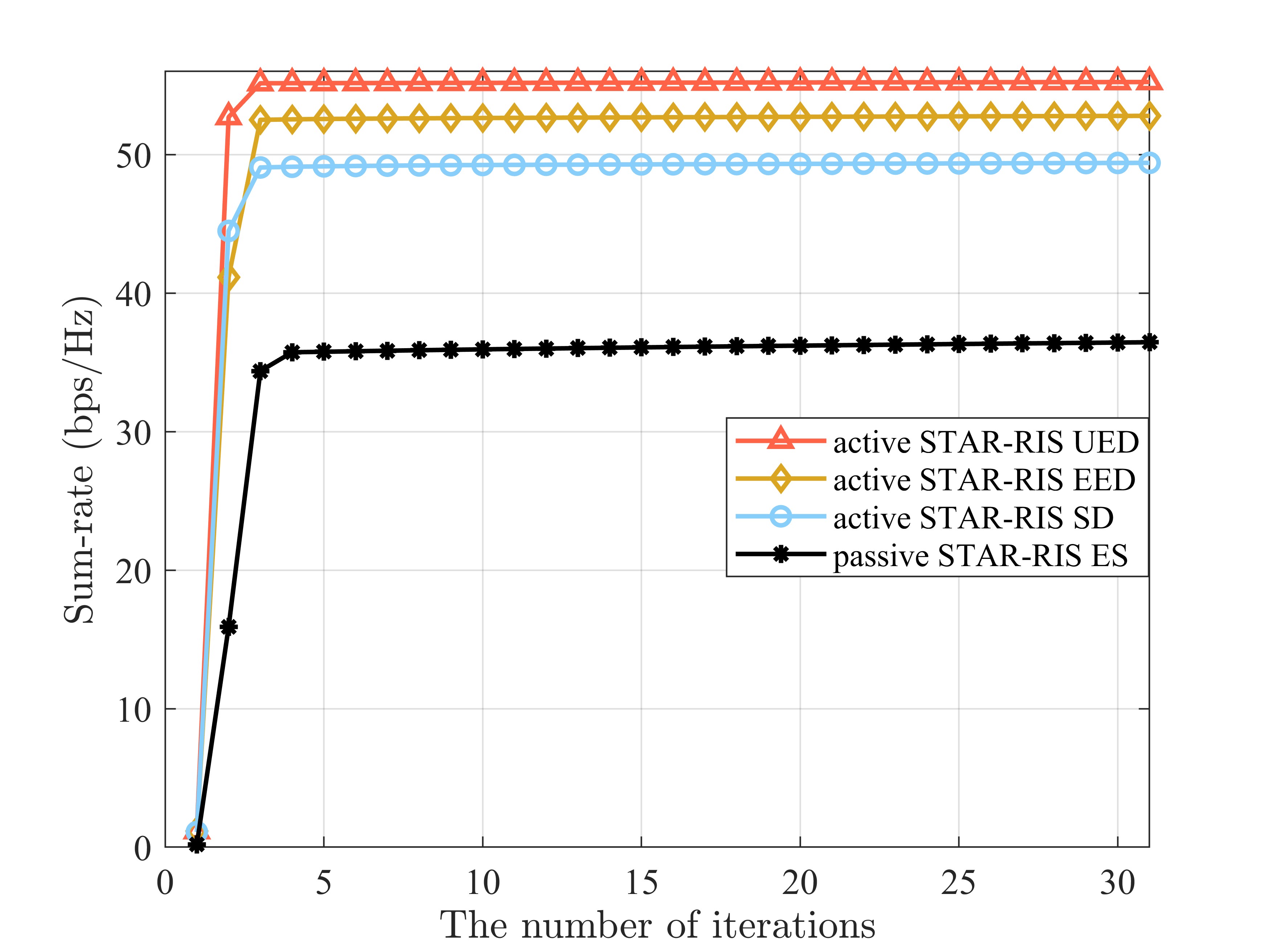}
	\caption{Achievable sum-rate versus the number of iterations.}
	\label{fig_4}
\end{figure}
\begin{figure}[!t]
	\centering
	\includegraphics[height=2.8in,width=3.4in]{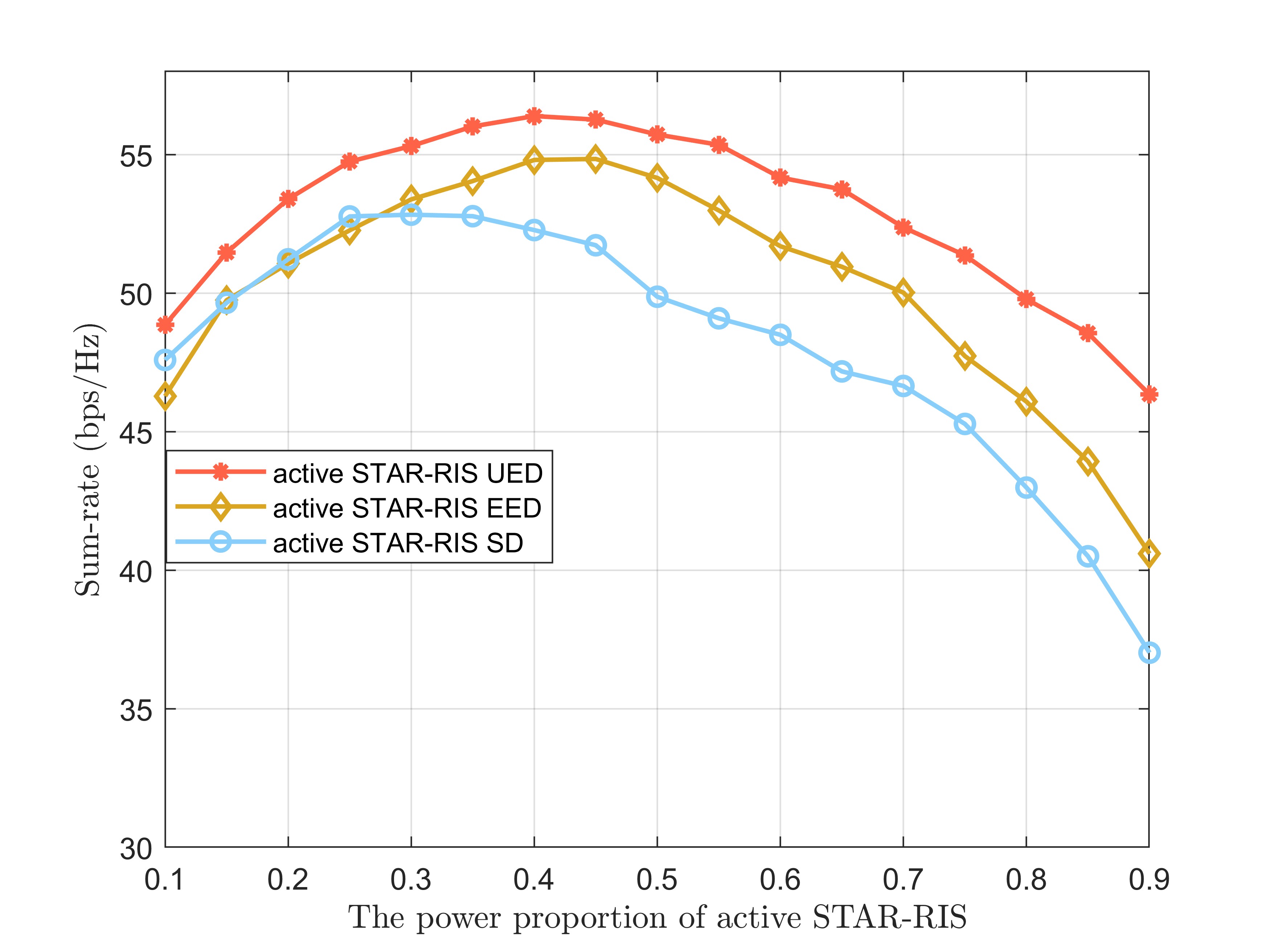}
	\caption{Achievable sum-rate versus the different power allocations between DFBS and active STAR-RIS.}
	\label{fig_5}
\end{figure}
\begin{figure}[!t]
	\centering
	\includegraphics[height=2.8in,width=3.4in]{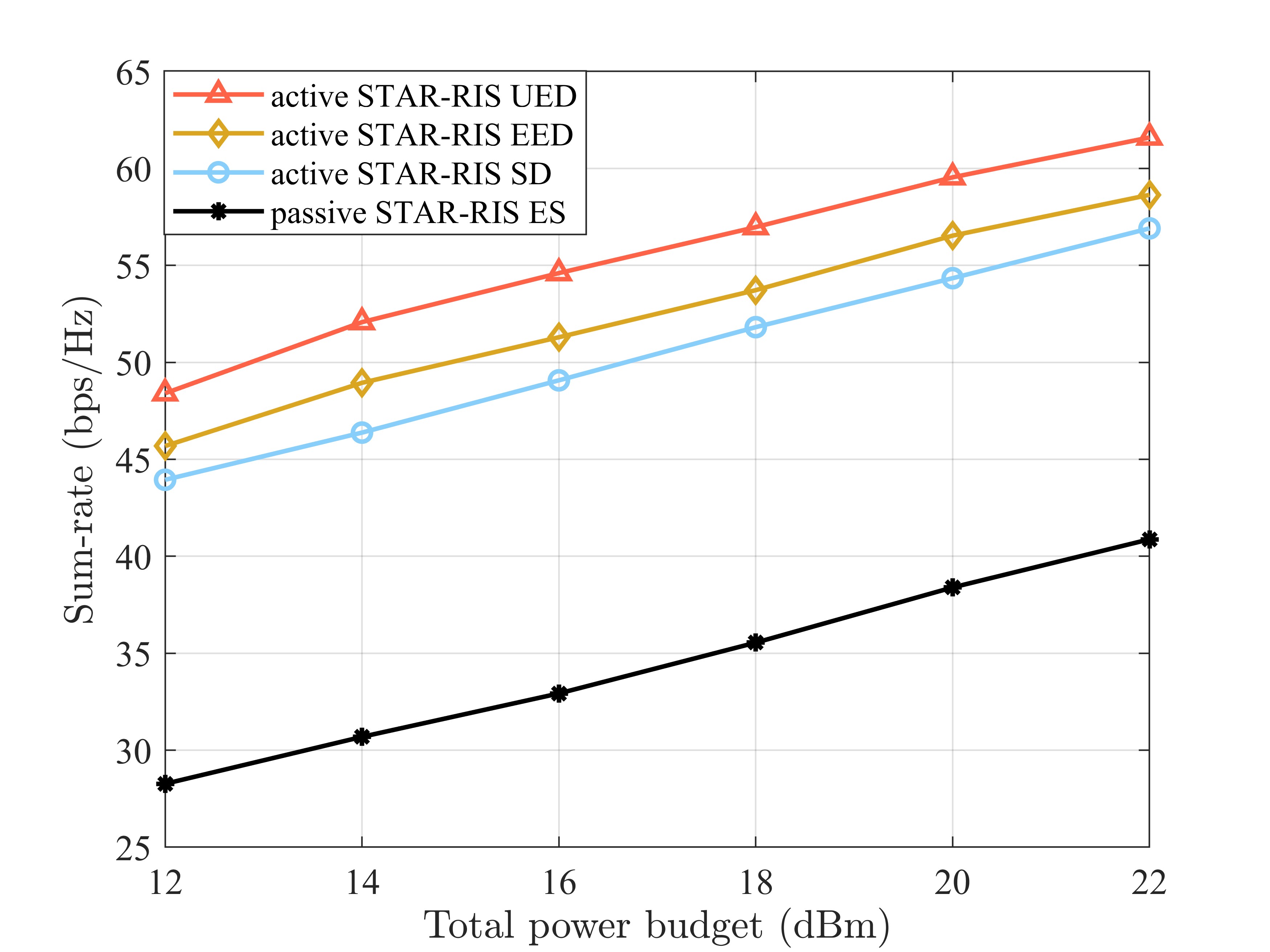}
	\caption{Achievable sum-rate versus total power budget $P_{\text{max}}^{\text{total}}$.}
	\label{fig_6}
\end{figure}
\begin{figure}[!t]
	\centering
	\includegraphics[height=2.8in,width=3.4in]{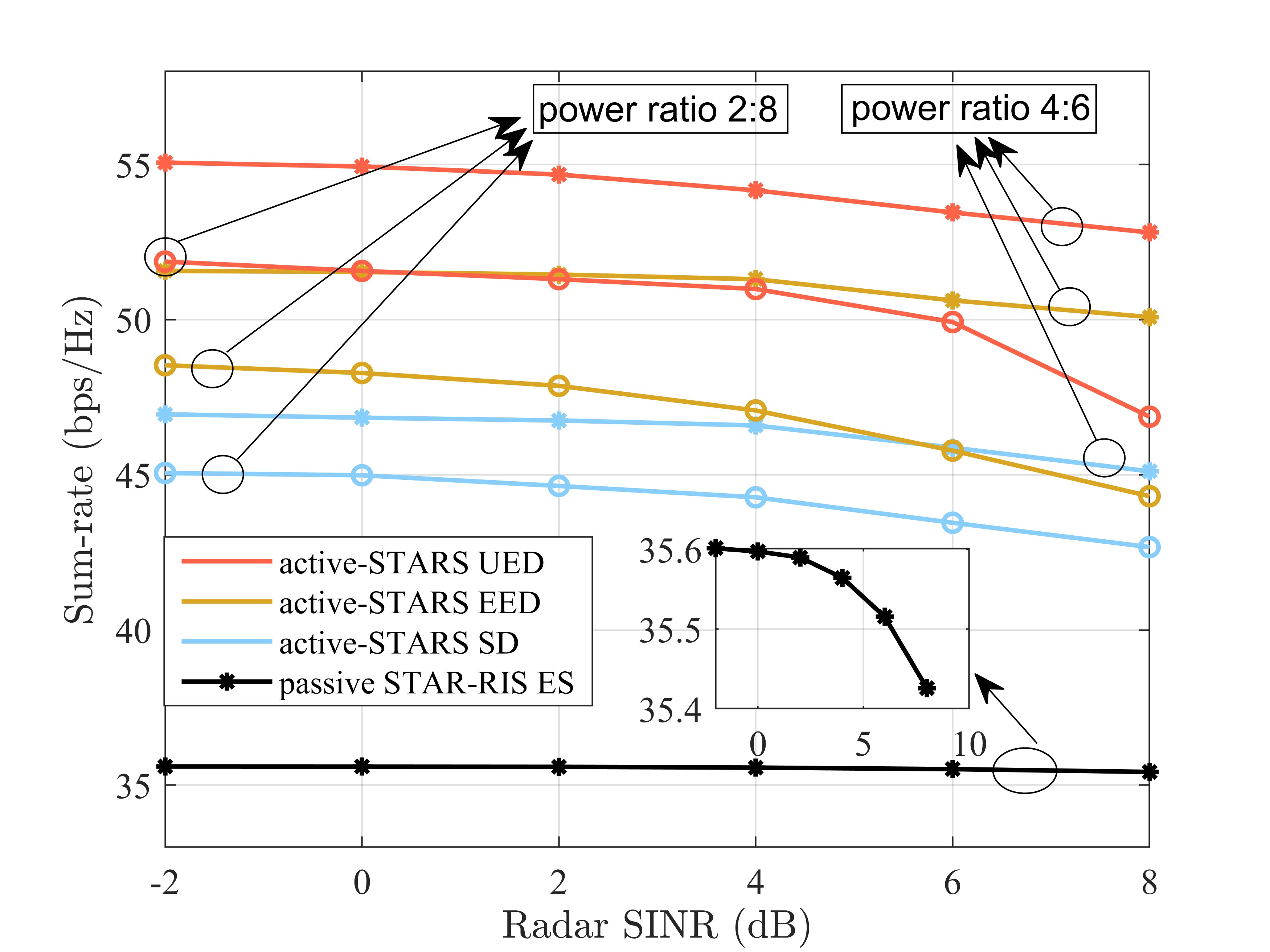}
	\caption{Achievable sum-rate versus Rader SINR $\Gamma_{t}$ with different power allocations between DFBS and active STAR-RIS.}
	\label{fig_7}
\end{figure}

Fig. \ref{fig_4} shows the convergence of the proposed algorithms. In order to ensure a fair comparison with the case of the passive STAR-RIS, we define the total power budget $P_{\text{max}}^{\text{total}}=P_{\text{max}}^{\text{B}}+P_{\text{max}}^{\text{R}}$. Thus, for passive scheme, the total power denotes the the DFBS transmit power, i.e., $P_{\text{max}}^{\text{total}}=P_{\text{max}}^{\text{B}}$. For our proposed schemes, we set the power budget at DFBS and active STAR-RIS to $P_{\text{max}}^{\text{B}}=0.4\times P_{\text{max}}^{\text{total}}$ and $P_{\text{max}}^{\text{R}}=0.6\times P_{\text{max}}^{\text{total}}$, respectively. Note that this kind of power allocation is just a case to verify the convergence and efficiency of our proposed algorithms. Actually, the achievable sum-rate will be different for different allocations, but this does not affect the effectiveness of our proposed algorithms. We notice in Fig. \ref{fig_4} that the sum-rate first experiences an upward trend and subsequently keeps stabilization as the number of iterations increases under all schemes, which validates the effectiveness of our proposed algorithms. In addition, one can observe that the sum-rate with UED mode is the highest, while the passive STAR-RIS owns the lowest sum-rate.

We plot Fig. \ref{fig_5} to show the relationship between achievable sum-rate and different power allocations between DFBS and active STAR-RIS. The number on $x$ axis, for example, 0.4 represents that under a certain total power budget $P_{\text{max}}^{\text{total}}$, the power allocated to active STAR-RIS is $P_{\text{max}}^{\text{R}}=0.4\times P_{\text{max}}^{\text{total}}$, and the power allocated to DFBS is $P_{\text{max}}^{\text{B}}=0.6 \times P_{\text{max}}^{\text{total}}$. One can observe that when the system's total power budget is given, different power allocations will lead to different performance, and there exists an optimal power allocation for each mode. Meanwhile, we can also find that the power allocation with middle ratio can obtain a better performance under UED and EED modes, because this ratio can not only satisfy the detection requirements of DFBS but also fully leverage the amplification characteristics of the active STAR-RIS and its ability to counteract the multiplicative fading effect. For SD mode, due to the low DoFs of active STAR-RIS elements, it needs more power consumption at the DFBS to overcome the double fading. Additionally, it can be observed that the performance of the UED mode is optimal, which once again confirms the UED mode owns the highest DoFs.

Fig. \ref{fig_6} shows the achievable sum-rate versus total power budget $P_{\text{max}}^{\text{total}}$. We can note that the sum-rate exhibits a gradual rise with the total power budget increasing under all schemes, which is easy to understand. Obviously, the UED scheme owns the highest sum-rate compared to other schemes. This can be explained as follows: When active STAR-RIS operates in UED mode, each element can be dynamically adjusted based on the given power of the active STAR-RIS. This allows active STAR-RIS to divide the incident signals upon each element flexibly into transmission and reflection parts with different amplitudes based on actual conditions, thus giving this mode the highest DoFs. Meanwhile, due to the multiplicative fading effect, the performance of the passive scheme is the worst compared to other schemes, demonstrating the effectiveness of deploying the active STAR-RIS in ISAC systems.

\begin{figure}[!t]
	\centering
	\includegraphics[height=2.8in,width=3.4in]{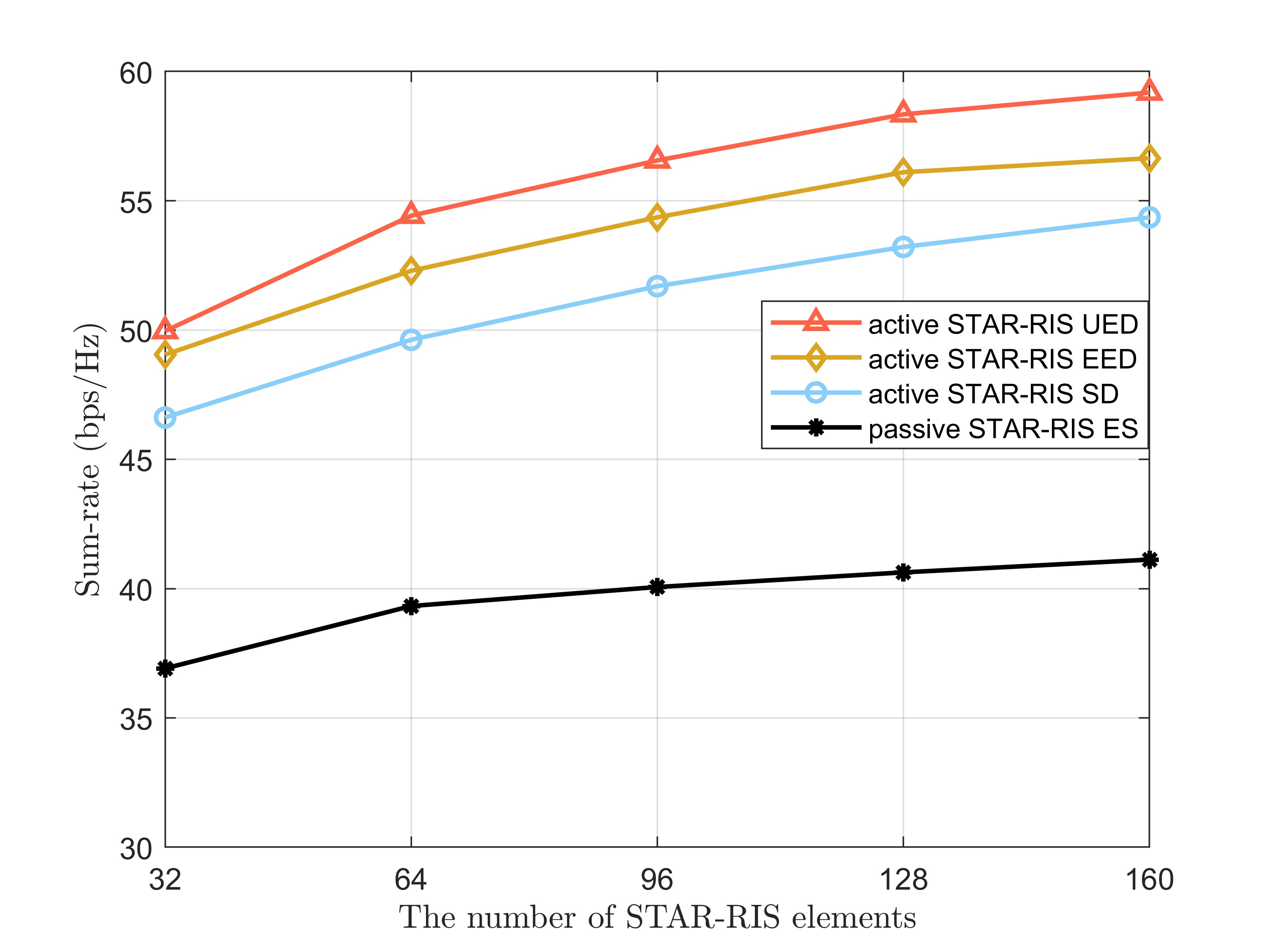}
	\caption{Achievable sum-rate versus active STAR-RIS elements $N$.}
	\label{fig_8}
\end{figure}

\begin{figure}[!t]
	\centering
	\includegraphics[height=2.8in,width=3.4in]{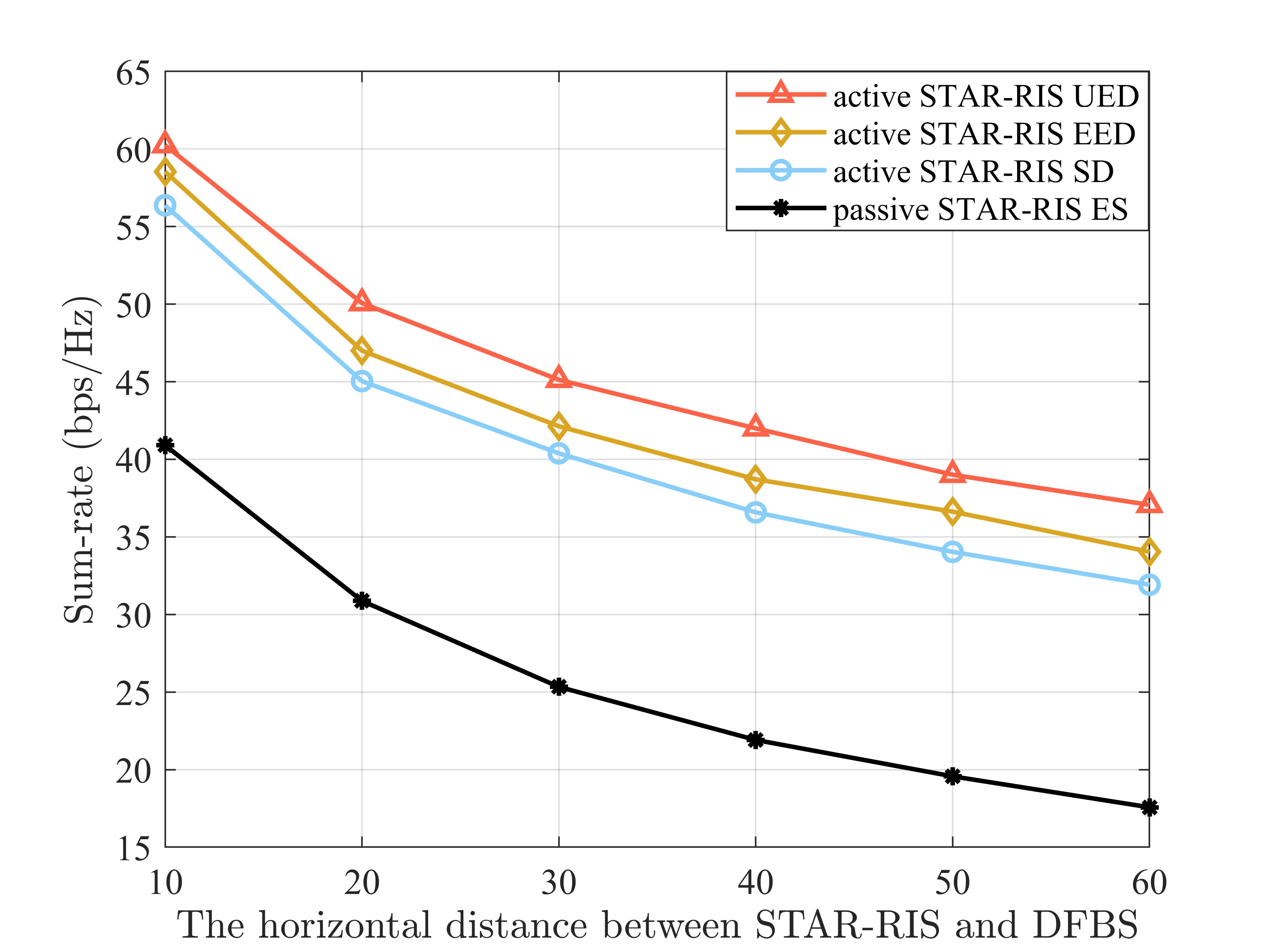}
	\caption{Achievable sum-rate versus the horizontal distance of the active STAR-RIS from the DFBS.}
	\label{fig_9}
\end{figure}
Fig. \ref{fig_7} shows the achievable sum rate versus radar SINR under different power allocations between DFBS and active STAR-RIS. The number ``4:6" indicates that with the given total power budget $P_{\text{max}}^{\text{total}}$, the power allocated to DFBS is $P_{\text{max}}^{\text{B}}=\times P_{\text{max}}^{\text{total}}$ and the power allocated to active STAR-RIS is $P_{\text{max}}^{\text{R}}=\times P_{\text{max}}^{\text{total}}$, and the same goes for ``2:8". As clearly seen, with the same radar SINR constraint, the ``4:6" allocation method obtained a better performance, consistent with our previous analysis. Moreover, one can find that the sum rate under all schemes decreases with radar SINR increases. This can be explained, as more power will be used for target sensing under a higher radar SINR, and thus, the power used for communication will be reduced, leading to a decrease in the sum-rate. This observation highlights the tradeoff between communication and sensing within the limited resources, and the realization of high sensing performance comes at the cost of diminished communication performance. In addition, the active STAR-RIS can always achieve a performance improvement compared to the passive one, demonstrating the significant role of deploying active STAR-RIS in ISAC systems and the effectiveness of the proposed algorithms.

Fig. \ref{fig_8} shows the sum-rate versus the number of STAR-RIS elements. It is obvious that the increase in the number of STAR-RIS elements leads to an enhancement in the sum-rate under all schemes. This is easy to understand, more STAR-RIS elements can obtain a higher beamforming gain, enhancing the sum-rate. One can also observe that the sum-rate improvement by the active schemes is much higher than that for the passive scheme. This indicates that, unlike passive scheme where all transmit power suffers from multiplicative fading of the reflected/transmitted links, the power of active STAR-RIS only experiences fading of the active STAR-RIS-user link, thus the power attenuation is much smaller and can effectively compensate for multiplicative fading effects. This also confirms that active STAR-RIS can achieve the same performance as the passive one with much fewer elements, which agrees with our analysis in Section I.

Finally, in order to investigate the influence of the STAR-RIS location on the sum-rate, we analyze the achievable sum-rate in relation to the horizontal distance between the STAR-RIS and the DFBS in Fig. \ref{fig_9}. In our considered scenario, it can be found that when the distance between DFBS and STAR-RIS is increased, the system is able to function normally. However, in practical systems, we need to consider the requirements of the actual situation. If the system's requirements for sum-rate are relatively low, then STAR-RIS deployment  can be far from DFBS. Conversely, if the requirements are high, it needs to be closer to DFBS. Taking our simulation as an example, if the required communication rate of the system cannot be lower than 45bps/Hz, then only when the distance between DFBS and STAR-RIS is less than 20m, the three modes of active STAR-RIS can be effective. In addition, as clearly observed, the sum-rate decreases as the distance between DFBS and STAR-RIS increases. This can be attributed to the increased path loss experienced by the DFBS-RIS link as the distance between STAR-RIS and DFBS increases. For passive STAR-RIS, it will be affected by significant path loss when the STAR-RIS is located at a considerable distance from the DFBS. However, in the case of active STAR-RIS, a portion of the total power budget is assigned to the active STAR-RIS that is only impacted by the slight path loss between the active STAR-RIS and user link, which can compensate for significant path loss between DFBS and STAR-RIS. Thus, the active STAR-RIS exhibits the capability to achieve a higher sum-rate compared to the passive STAR-RIS.

\section{Conclusion}
In this paper, we have investigated an ISAC system that utilizes active STAR-RIS to overcome the challenges of multiplicative fading and limited coverage. Meanwhile, we have considered multiple targets and multiple users scenario and assumed the existence of SI. In particular, we have formulated an optimization problem with the goal of maximizing the achievable communication sum-rate, while also considering the requirement of sensing, the power budget and the active STAR-RIS hardware constraints. An AO algorithm have been proposed to efficiently solve the formulated problem by decomposing it into several tractable subproblems. Through fixing variables, these subproblems can be easily tackled with standard convex optimization algorithms and state-of-the-art algorithms. Simulation results have shown the potential of deploying active STAR-RIS in ISAC system and the efficiency of the proposed joint beamforming optimization algorithms. In future, we will explore the scenario based on imperfect channel state information to enhance the practicality of our research.

\begin{IEEEbiography}[{\includegraphics[width=1in,height=1.25in,clip,keepaspectratio]{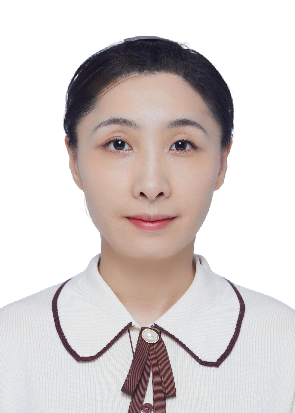}}]{Shuang Zhang} received the M.S. degree in Communication and Information System from Xidian University, Xi'an, China, in 2017. She is currently pursuing the Ph.D. degree in information and communication engineering with Zhengzhou University, Zhengzhou, China. Her research interests include integrated sensing and communication (ISAC) and reconfigurable intelligent surface (RIS). 
\end{IEEEbiography}
\begin{IEEEbiography}[{\includegraphics[width=1in,height=1.25in,clip,keepaspectratio]{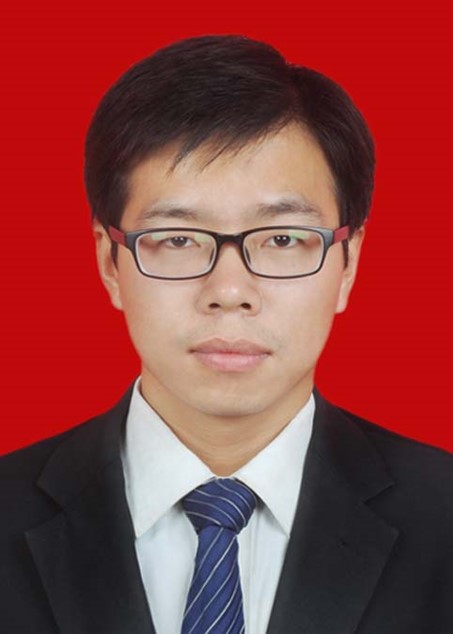}}]{Wanming Hao} (Senior Member, IEEE) received the Ph.D. degree from School of Electrical and Electronic Engineering, Kyushu University, Japan, in 2018. He worked as a Research Fellow at the 5G Innovation Center, Institute of Communication Systems, University of Surrey, U.K. Now, he is	an Associate Professor at the School of Electrical and Information Engineering, Zhengzhou University, China. His research interests include millimeterwave and RIS. He is the editor of IEEE O-JCS.
\end{IEEEbiography}
\begin{IEEEbiography}[{\includegraphics[width=1in,height=1.25in,clip,keepaspectratio]{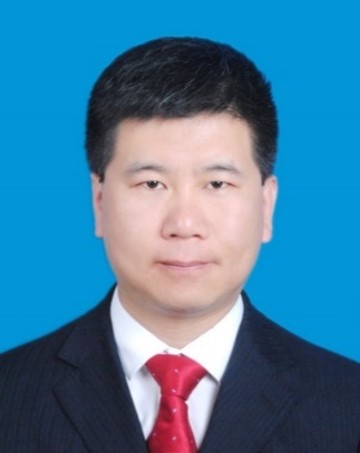}}]{Gangcan Sun} received his Ph.D. degree in Communication and Information System from Beijing Institute of Technology, China, in 2009. He was promoted to an associate professor in School of Information Engineering, Zhengzhou University, China, in 2013, and now he has a full professor. His research interests include communication signal processing, communication modulation recognition and blind estimation of key parameters of communication signals.
\end{IEEEbiography}
\begin{IEEEbiography}[{\includegraphics[width=1in,height=1.25in,clip,keepaspectratio]{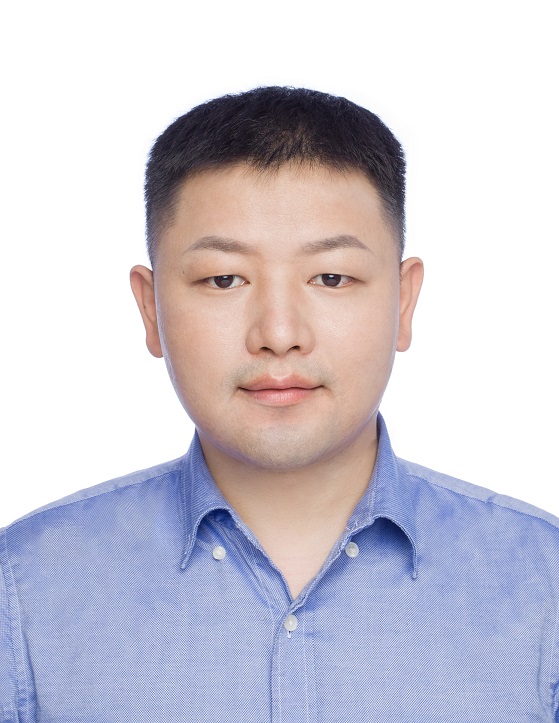}}]{Chongwen Huang} (Member, IEEE) received his B.Sc. degree in 2010 from Nankai University, and the M.Sc. degree from the University of Electronic Science and Technology of China in 2013, and PhD degree from Singapore University of Technology and Design (SUTD) in 2019. From Oct. 2019 to Sep. 2020, he is a Postdoc in SUTD. Since Sep. 2020, he joined into Zhejiang University as a tenure-track young professor. Dr. Huang is the recipient of 2021 IEEE Marconi Prize Paper Award, 2023 IEEE Fred W. Ellersick Prize Paper Award and 2021 IEEE ComSoc Asia-Pacific Outstanding Young Researcher Award. He has served as an Editor of IEEE Communications Letter, Elsevier Signal Processing, EURASIP Journal on Wireless Communications and Networking and Physical Communication since 2021. His main research interests are focused on Holographic MIMO Surface/Reconfigurable Intelligent Surface, B5G/6G Wireless Communications, mmWave/THz Communications, Deep Learning technologies for Wireless communications, etc. 
\end{IEEEbiography}
\begin{IEEEbiography}[{\includegraphics[width=1in,height=1.25in,clip,keepaspectratio]{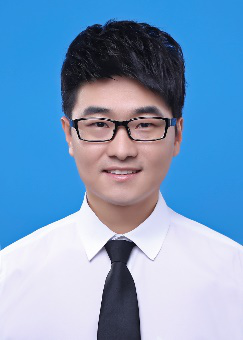}}]{Zhengyu Zhu} (Senior Member, IEEE) received the Ph.D. degree in information engineering from Zhengzhou University, Zhengzhou, China, in 2017. He is currently an Associate Professor with Zhengzhou University. From October 2013 to October 2015, he visited the Communication and Intelligent System Laboratory, Korea University, Seoul, South Korea, to conduct a collaborative research as a Visiting Student. His research interests include in formation theory and signal processing for wireless communications, such as B5G/6G, Intelligent reflecting surface, the Internet of Things, machine learning, millimeter wave communication, UAV communication, physical layer security, convex optimization techniques, and energy harvesting communication systems. He was an Associate Editor for the Journal of Communications and Networks, the Wireless Communications and Mobile Computing, and the Physical Communications from 2021.
\end{IEEEbiography}
\begin{IEEEbiography}[{\includegraphics[width=1in,height=1.25in,clip,keepaspectratio]{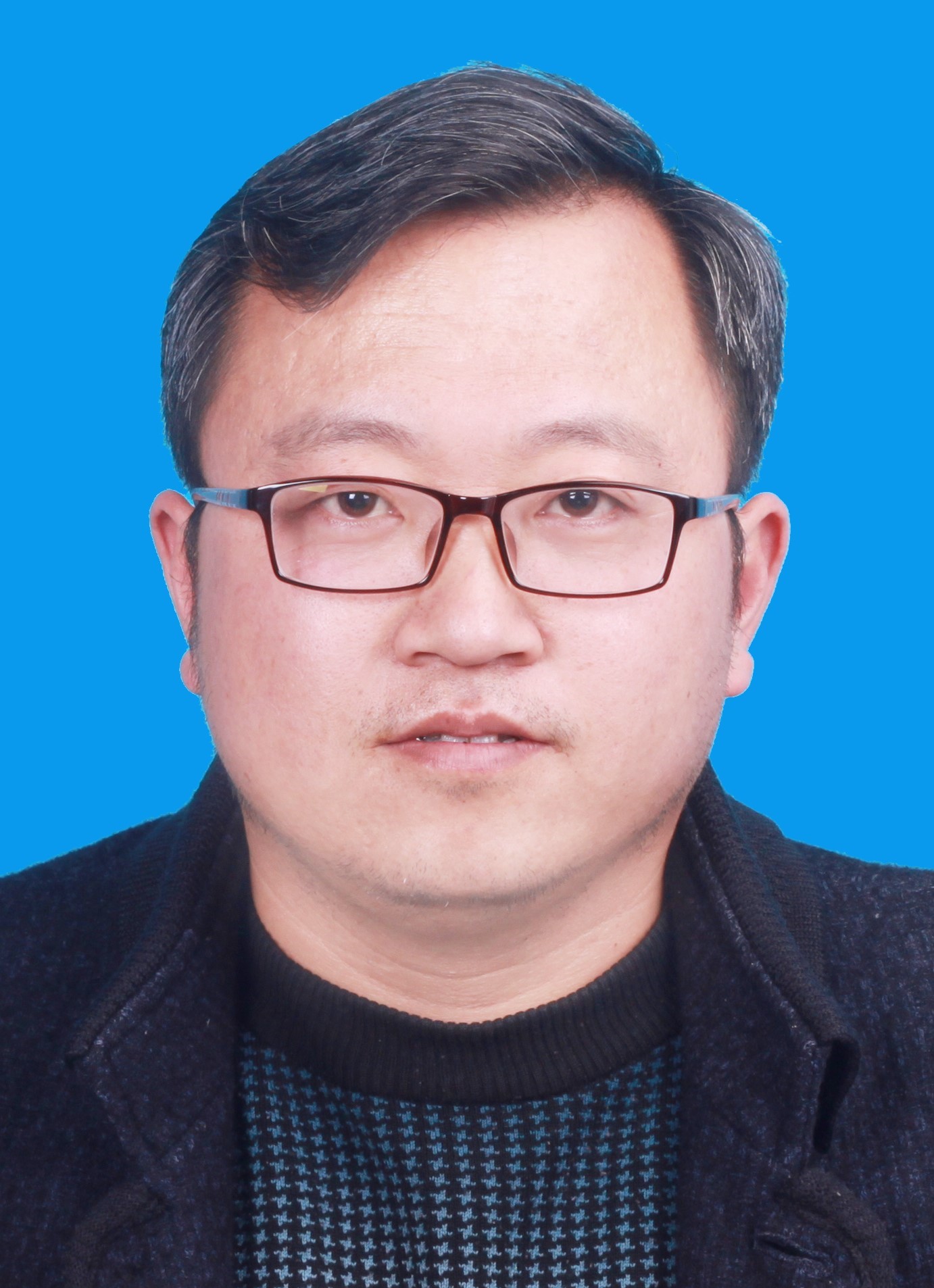}}]{Xingwang Li} (Senior Member, IEEE) received his M.Sc. and Ph.D. degrees from University of Electronic Science and Technology of China and Beijing University of Posts and Telecommunications in 2010 and 2015. From 2010 to 2012, he was working with Comba Telecom Ltd. in Guangzhou China, as an engineer. He spent one year from 2017 to 2018 as a visiting scholar at Queen’s University Belfast, Belfast, UK. He is currently an Associated Professor with the School of Physics and Electronic Information Engineering, Henan Polytechnic University, Jiaozuo China. His research interests span wireless communication, intelligent transport system, artificial intelligence, Internet of things. He was a recipient of exemplary Reviewer for IEEE Transactions on Communications and Journal of Electronics \& Information Technology in 2022. He is on the editorial board of IEEE Transactions on Intelligent Transportation Systems, IEEE Transactions on Vehicular Technology, IEEE Systems Journal, IEEE Sensors Journal, Physical Communication, etc. He has serviced as the Guest Editor for the special issue on Integrated Sensing and Communications (ISAC) for 6G IoE of IEEE Internet of Things Journal, Computational Intelligence and Advanced Learning for Next-Generation Industrial IoT of IEEE Transactions on Network Science and Engineering, “AI driven Internet of Medical Things for Smart Healthcare Applications: Challenges, and Future Trends” of the IEEE Journal of Biomedical and Health Informatics, etc.
\end{IEEEbiography}
\begin{IEEEbiography}[{\includegraphics[width=1in,height=1.25in,clip,keepaspectratio]{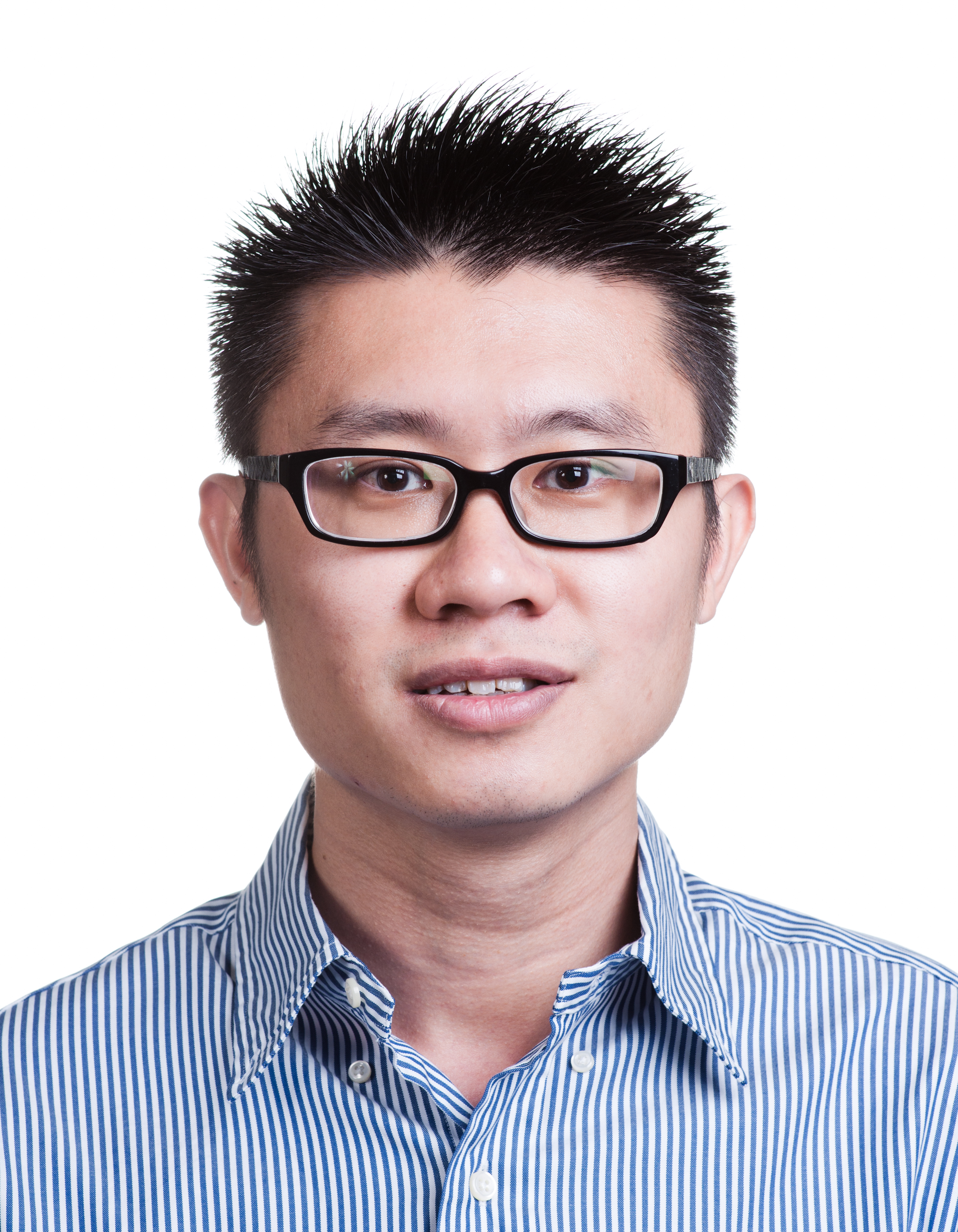}}]{Chau Yuen} (Fellow, IEEE) received the B.Eng. and Ph.D. degrees from Nanyang Technological University, Singapore, in 2000 and 2004, respectively. He was a Post-Doctoral Fellow with Lucent Technologies Bell Labs, Murray Hill, in 2005. From 2006 to 2010, he was with the Institute for Infocomm Research, Singapore. From 2010 to 2023, he was with the Engineering Product Development Pillar, Singapore University of Technology and Design. Since 2023, he has been with the School of Electrical and Electronic Engineering, Nanyang Technological University, currently he is Provost’s Chair in Wireless Communications, and Assistant Dean in Graduate College.

Dr. Yuen received IEEE Communications Society Leonard G. Abraham Prize (2024), IEEE Communications Society Best Tutorial Paper Award (2024), IEEE Communications Society Fred W. Ellersick Prize (2023), IEEE Marconi Prize Paper Award in Wireless Communications (2021), IEEE APB Outstanding Paper Award (2023), and EURASIP Best Paper Award for JOURNAL ON WIRELESS COMMUNICATIONS AND NETWORKING (2021).

Dr. Yuen current serves as an Editor-in-Chief for Springer Nature Computer Science, Editor for IEEE TRANSACTIONS ON VEHICULAR TECHNOLOGY, IEEE SYSTEM JOURNAL, and IEEE TRANSACTIONS ON NETWORK SCIENCE AND ENGINEERING, where he was awarded as IEEE TNSE Excellent Editor Award and Top Associate Editor for TVT from 2009 to 2015. He also served as the guest editor for several special issues, including IEEE JOURNAL ON SELECTED AREAS IN COMMUNICATIONS, IEEE WIRELESS COMMUNICATIONS MAGAZINE, IEEE
COMMUNICATIONS MAGAZINE, IEEE VEHICULAR TECHNOLOGY MAGAZINE, IEEE TRANSACTIONS ON COGNITIVE COMMUNICATIONS AND NETWORKING, and ELSEVIER APPLIED ENERGY. 

He is a Distinguished Lecturer of IEEE Vehicular Technology Society, Top 2$\%$ Scientists by Stanford University, and also a Highly Cited Researcher by Clarivate Web of Science. He has 4 US patents and published over 400 research papers at international journals.
\end{IEEEbiography}

\end{document}